\documentclass[twocolumn]{IEEEtran} 

\def\BibTeX{{\rm B\kern-.05em{\sc i\kern-.025em b}\kern-.08em
    T\kern-.1667em\lower.7ex\hbox{E}\kern-.125emX}}

\begin{document}

\title{
 Combinatorial
 Evolution and Forecasting
 of Communication
  Protocol ZigBee}


%
%
%
%
%
%
%

\author{Mark Sh. Levin, {\it Member, IEEE}
\thanks{Mark Sh. Levin is with
 Inst. for Inform. Transmission Problems,
 Russian Acad. of Sci.,
  Moscow 127994, Russia
  (e-mail: mslevin@acm.org).},
  Aliaksei Andrushevich
\thanks{Aliaksei Andrushevich,
 Rolf Kistler,
 Alexander Klapproth
 are with
 CEESAR-iHomeLab,
 Lucerne Univ. of Appl. Sci.,
  CH-6048 Horw, Switzerland
  (e-mail: aliaksei.andrushevich@hslu.ch,
  rolf.kistler@hslu.ch,
  alexander.klapproth@hslu.ch).},
  Rolf Kistler
,
  Alexander Klapproth
%
  }


\maketitle

\begin{abstract}
 The article addresses
 combinatorial evolution
 and forecasting
 of communication protocol for
 wireless sensor networks (ZigBee).
 Morphological tree structure (a version of and-or tree)
 is used as a hierarchical model for the protocol.
 Three generations of ZigBee protocol are examined.
 A set of protocol change operations is
 generated and described.
 The change operations are used as items
 for forecasting based on combinatorial  problems
 (e.g., clustering, knapsack problem, multiple choice knapsack problem).
 Two kinds of preliminary forecasts for the examined communication protocol are considered:
 (i) direct expert
 (expert judgment)
   based forecast,
 (ii) computation of the forecast(s)
 (usage of multicriteria decision making and combinatorial optimization
 problems).
 Finally, aggregation of the obtained preliminary forecasts
 is considered
 (two aggregation strategies are used).
\end{abstract}


\newcounter{cms}
\setlength{\unitlength}{1mm}


\section{Introduction}

 The significance of systems evolution/development and forecasting
 is increasing
 (e.g.,
 \cite{ayr69},
 \cite{dek07},
 \cite{sahal81}).
 In the case of hierarchical modular systems,
 combinatorial approaches to systems evolution and forecasting
 were proposed
 in
 (\cite{lev02},  \cite{lev06},  \cite{lev11agg}).
 The approaches are based on hierarchical system modeling and usage of
 multicriteria decision making and combinatorial optimization
 problems.
 Some applied examples of combinatorial evolution and forecasting
 (e.g., electronic equipment, standard for transmission of
 multimedia data)
 have been described in
 (\cite{lev06},  \cite{lev11agg}, \cite{lev10}).
 In recent years,  wireless sensor networks
 are widely used in many domains
 (e.g.,
  \cite{ak02},
  \cite{bar07},
  \cite{chen11},
   \cite{dek07},
   \cite{hac03},
  \cite{karl07},
 \cite{wh07},  \cite{yick08}).
 Here many research works are targeted to
 analysis and synthesis (e.g., optimization)
 of communication protocols for wireless sensor networks
 (e.g.,
  \cite{bar07}, \cite{karl07}, \cite{tate12}).
 In the paper,
 combinatorial evolution and forecasting of  protocol ZigBee
 for wireless sensor
 networks
 (\cite{bar07}, \cite{karl07}, \cite{wh07})
  are considered.
 A morphological tree structure (a version of {\it and-or tree})
 is used as a hierarchical modular model for the protocol.
 Three generations of ZigBee protocol are examined:
 (1) ZigBee 2004 \(S_{1}\),
 (2) ZigBee 2006 \(S_{2}\), and
 (3) ZigBee PRO \(S_{3}\).
 A set of protocol change operations
 (between protocol generations above)
  is
 generated and described.
 The change operations are used as items
 for forecasting based on combinatorial problems
 (e.g., clustering, knapsack problem,
 multiple choice knapsack problem,
 multicriteria ranking).
 Two kinds of preliminary forecasts for the examined communication protocol are considered:
 (i) direct expert based forecast
 ~ \(\widetilde{S_{4}}\)  (\(\widetilde{\Phi}\)),
 (ii) two computed forecasts
 (usage of multicriteria decision making and combinatorial optimization
 problems):
 \(\widehat{\Phi}\)
 and
 \(\overline{\Phi}\).
 Further, the obtained three preliminary forecasts above
 are aggregated to build  resultant
 forecasts:
  \(\Theta^{I}\)
 (aggregation strategy I)
   and \(\Theta^{II}\)
 (aggregation strategy II).

 A flowchart of the article is as follows:
 (1) designing a general tree-like model
 of ZigBee protocol;
 (2) description of three protocol generations (including their
 structures and components);
 (3) expert judgement to obtain a direct expert based (preliminary) forecast;
 (4) extraction of changes between neighbor protocol generations;
 (5) generation of an integrated set of basic change operations;
 (6) evaluation of change operations upon criteria;
 (7) solving of combinatorial problems (ranking, clustering) and
 forecasting (e.g.,
 multicriteria choice knapsack problem)
 to obtain computed preliminary forecasts; and
 (8) aggregation of the obtained
 preliminary forecasts
 to build a resultant aggregated forecast(s)
 (two aggregation strategies are used).
%
 The paper is based on preliminary materials:
 (i) conference paper
 (combinatorial evolution, preliminary forecasts
 \cite{levand10})
 and
 (ii) electronic preprint
 (aggregation approaches and aggregation example
  \cite{lev11agg}).

\section{General Scheme}

A general framework of combinatorial evolution and forecasting
 for modular systems
 in case of three system generations is depicted in Fig. 1
 (\cite{lev06},  \cite{lev11}).

\begin{center}
\begin{picture}(80,76)
\put(13,0){\makebox(0,0)[bl]{Fig. 1. General framework
(\cite{lev06},
 \cite{lev11})
 }}

\put(40,71){\oval(38,6)} \put(40,71){\oval(37,5)}

\put(24,69.5){\makebox(0,0)[bl]{Aggregated forecast(s)}}


\put(40,64){\vector(0,1){4}}

\put(6,58){\line(1,0){68}} \put(6,64){\line(1,0){68}}
\put(6,58){\line(0,1){06}} \put(74,58){\line(0,1){06}}

\put(6.5,58.5){\line(1,0){67}} \put(6.5,63.5){\line(1,0){67}}
\put(6.5,58.5){\line(0,1){05}} \put(73.5,58.5){\line(0,1){05}}

\put(25,59.2){\makebox(0,8)[bl]{Aggregation process}}


\put(15,54){\vector(0,1){4}} \put(65,54){\vector(0,1){4}}


\put(07.5,50){\makebox(0,8)[bl]{Forecast \(1\)
}}

\put(6,48){\line(1,0){18}} \put(6,54){\line(1,0){18}}
\put(6,48){\line(0,1){06}} \put(24,48){\line(0,1){06}}

\put(6.5,48){\line(0,1){06}} \put(23.5,48){\line(0,1){06}}


\put(36,50.5){\makebox(0,8)[bl]{{\bf .~ .~ . }}}


\put(57.5,50){\makebox(0,8)[bl]{Forecast \(n\)
}}

\put(56,48){\line(1,0){18}} \put(56,54){\line(1,0){18}}
\put(56,48){\line(0,1){06}} \put(74,48){\line(0,1){06}}

\put(56.5,48){\line(0,1){06}} \put(73.5,48){\line(0,1){06}}


\put(15,44){\vector(0,1){4}} \put(65,44){\vector(0,1){4}}


\put(02,38){\line(1,0){76}} \put(02,44){\line(1,0){76}}
\put(02,38){\line(0,1){06}} \put(78,38){\line(0,1){06}}

\put(03,39.5){\makebox(0,0)[bl]{Forecasting (optimization models,
expert judgment)}}


\put(32,33.5){\vector(0,1){4}} \put(40,33.5){\vector(0,1){4}}
\put(48,33.5){\vector(0,1){4}}

\put(40,31){\oval(50,5)}

\put(20,29){\makebox(0,0)[bl]{Set of change operations}}


\put(25,24.5){\vector(0,1){4}}

\put(17,14){\vector(1,1){4}} \put(29,18){\vector(1,-1){4}}

\put(25,22){\oval(14,5)}

\put(19,20){\makebox(0,0)[bl]{Changes}}


\put(55,24.5){\vector(0,1){4}}

\put(47,14){\vector(1,1){4}} \put(59,18){\vector(1,-1){4}}

\put(55,22){\oval(14,5)}

\put(49,20){\makebox(0,0)[bl]{Changes}}


\put(00,06){\line(1,0){20}} \put(00,06){\line(1,2){10}}
\put(20,06){\line(-1,2){10}}

\put(5,11){\makebox(0,0)[bl]{System}}
\put(09,07.5){\makebox(0,0)[bl]{\(1\)}}


\put(30,06){\line(1,0){20}} \put(30,06){\line(1,2){10}}
\put(50,06){\line(-1,2){10}}

\put(35,11){\makebox(0,0)[bl]{System}}
\put(39,07.5){\makebox(0,0)[bl]{\(2\)}}


\put(60,06){\line(1,0){20}} \put(60,06){\line(1,2){10}}
\put(80,06){\line(-1,2){10}}

\put(65,11){\makebox(0,0)[bl]{System}}
\put(69,07.5){\makebox(0,0)[bl]{\(3\)}}

\end{picture}
\end{center}

 Here  a hierarchical multi-layer system model
  ``morphological tree'' is used
 (\cite{lev06}, \cite{lev11agg}):
 (i) tree-like system model,
 (ii) set of leaf nodes as basic system parts/components,
 (iii) sets of design alternatives (DAs) for each leaf node,
 (iv) DAs rankings (i.e., ordinal priorities), and
 (v) compatibility estimates between DAs for different leaf nodes.
 This ``morphological tree'' model
 is a version of ``and-or tree''.
 In this paper, a simplified version of our ``morphological tree''
 is used
 without estimates of DAs and their compatibility.
 Our scheme for evolution and forecasting of ZigBee protocol
 is depicted in Fig. 2.

\begin{center}
\begin{picture}(72,54)
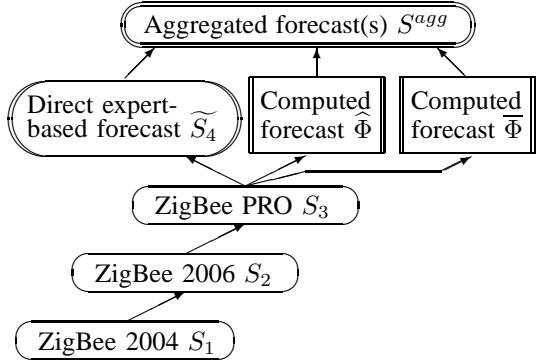
\put(00,00){\makebox(0,0)[bl]{Fig. 2. Evolution, forecasting
 for ZigBee protocol}}


\put(15,08){\oval(29,5)}

\put(03.5,06){\makebox(0,8)[bl]{ZigBee 2004 \(S_{1}\)}}

\put(15,10.5){\vector(2,1){8}}


\put(23,17){\oval(30,5)}

\put(11,15){\makebox(0,8)[bl]{ZigBee 2006 \(S_{2}\)}}

\put(23,19.5){\vector(2,1){8}}


\put(31,26){\oval(30,5)}

\put(31,28.5){\vector(2,1){8}} \put(31,28.5){\vector(-2,1){8}}

\put(19,24){\makebox(0,8)[bl]{ZigBee PRO \(S_{3}\)}}

\put(31,28.5){\line(4,1){8}}

\put(39,30.5){\line(1,0){18}}

\put(57,30.5){\vector(2,1){4}}


\put(15,37.5){\oval(30,10)}

\put(15,37.5){\oval(31,10)}

\put(2,37.5){\makebox(0,8)[bl]{Direct expert-}}

\put(2,34.5){\makebox(0,8)[bl]{based forecast
\(\widetilde{S_{4}}\)}}


\put(33,38){\makebox(0,8)[bl]{Computed }}

\put(33,34.9){\makebox(0,8)[bl]{forecast
 \(\widehat{\Phi}\)
 }}

\put(31.5,33){\line(1,0){18}} \put(31.5,43){\line(1,0){18}}
\put(31.5,33){\line(0,1){10}} \put(49.5,33){\line(0,1){10}}

\put(32,33){\line(0,1){10}} \put(49,33){\line(0,1){10}}


\put(53,38){\makebox(0,8)[bl]{Computed }}

\put(53,34.9){\makebox(0,8)[bl]{forecast \(\overline{\Phi}\)
 }}

\put(51.5,33){\line(1,0){18}} \put(51.5,43){\line(1,0){18}}
\put(51.5,33){\line(0,1){10}} \put(69.5,33){\line(0,1){10}}

\put(52,33){\line(0,1){10}} \put(69,33){\line(0,1){10}}


\put(60.5,43){\vector(-1,1){4}} \put(40.5,43){\vector(0,1){4}}

\put(15,43){\vector(1,1){4}}

\put(38,50){\oval(46,05)}

\put(38,50){\oval(47,06)}

\put(17.5,48.2){\makebox(0,8)[bl]{Aggregated forecast(s)
\(S^{agg}\)}}

\end{picture}
\end{center}

\section{Description of Protocol Generations}

 Let us consider hierarchical structures
 (as {\it and/or trees})
 for three basic versions of ZigBee protocols.
 The structure of generation 1
 ZigBee 2004  (\(S_{1}\)) is the following:

 {\it 1.} Interference avoidance \(A\):~
 \(A_{1}\) (PAN coordinator selects best available RF channel/Network ID at
 startup time).

 {\it 2.} Automated/distributed address management \(B\):~
 \(B_{1}\)
 (Device addresses automatically assigned using a hierarchical,
 distributed scheme).

 {\it 3.} Centralized data collection \(C\):~
       {\it 3.1.} Low-overhead data collection by ZigBee Coordinator
       \(G\):~
       \(G_{1}\) (Fully supported),
       {\it 3.2.} Low-overhead data collection by other devices
       \(H\):~
       \(H_{1}\) (Under special circumstances).

  {\it 4.} Network scalability \(D\):~
  \(D_{1}\) (Network scales up to the limits of
 the addressing algorithm. Typically, networks with tens to
 hundreds of devices are supported).

 {\it 5.} Message size \(E\):~
 \(E_{1}\)
 (\(<\)100 bytes. Exact size depends on services employed, such as
 security).

 {\it 6.} Robust mesh networking \(F\):~
 \(F_{1}\) (Fault tolerant routing
 algorithms respond to changes in the network and in the RF
 environment).

 The structure of generation 2
 ZigBee 2006  (\(S_{2}\)) is the following:

 {\it 1.} Interference avoidance \(A\):~
 \(A_{1}\)
 (PAN coordinator selects best
 available RF channel/Network ID at startup time).

 {\it 2.} Automated/distributed address management \(B\):~
 \(B_{1}\) (Device addresses
 automatically assigned using a hierarchical, distributed scheme).

 {\it 3.} Group addressing \(I\):~
 \(I_{1}\) (Devices can be assigned to groups, and
 whole groups can be addressed with a single frame).

 {\it 4.} Centralized data collection \(C\):~
   {\it 4.1.} Low-overhead data collection by ZigBee Coordinator
       \(G\):~
       \(G_{1}\) (Fully supported),
   {\it 4.2.} Low-overhead data collection by other devices
   \(H\):~
       \(H_{1}\) (Under special circumstances).

 {\it 5.} Network scalability \(D\):~
 \(D_{1}\) (Network scales up to the limits of
 the addressing algorithm. Typically, networks with tens to
 hundreds of devices are supported).

 {\it 6.} Message size \(E\):~
 \(E_{1}\) (\(<\)100 bytes. Exact size depends on
 services employed, such as security).

 {\it 7.} Standardized commissioning \(K\):~
 \(K_{1}\)
 (Standardized startup procedure and attributes
 support the use of commissioning tools in a multi-vendor
 environment).

 {\it 8.} Robust mesh networking \(F\):~
 \(F_{1}\) (Fault tolerant routing
 algorithms respond to changes in the network and in the RF
 environment).

 {\it 9.} Cluster Library support \(L\):~
 \(L_{1}\) (The ZigBee Cluster
 Library, as an adjunct to the stack, standardizes application
 behavior across profiles and provides an invaluable resource for
 profile developers).
 The structure of generation 3
 ZigBee PRO  (\(S_{3}\)) is the following:

   {\it 1.} Interference avoidance \(A'\):~
     {\it 1.1.} Startup Procedure of Channel Acquisition \(M\):~
     \(M_{1}\)
     (PAN coordinator selects best available RF channel/Network ID at startup
     time),
     {\it 1.2.} Channel Hopping \(N\):~
     \(N_{1}\) (Ongoing interference detection and adoption
     of a new operating RF channel and/or Network ID).

 {\it 2.} Automated/distributed address management \(B\):~
 \(B_{2}\)
  (Device addresses automatically assigned using a stochastic scheme.)

 {\it 3.} Group addressing \(I\):~
 \(I_{1}\) (Devices can be assigned to groups, and
 whole groups can be addressed with a single frame).

 {\it 4.} Centralized data collection \(C'\):~
       {\it 4.1.} Low-overhead data collection by ZigBee Coordinator
       \(G\):~
       \(G_{1}\) (Fully supported),
       {\it 4.2} Low-overhead data collection by other devices
       \(H\):~
       \(H_{1}\) (Under special circumstances),
       {\it 4.3.} Many-to-one routing \(Q\):~
       \(Q_{1}\) (Whole network discovers the aggregator in one pass), and
       {\it 4.4.} Source routing \(P\):~
       \(P_{1}\)
        (Aggregator responds to all senders in an economical manner).

 {\it 5.} Network scalability \(D\):~
 \(D_{2}\) (An addressing algorithm that
 relaxes the limits on network size. Networks with hundreds to
 thousands of devices are supported).

 {\it 6.} Message size \(E\):~
  \(E_{2}\) (Large messages, up to the buffer capacity of the sending and receiving
 devices, are supported using Fragmentation and Reassembly).

 {\it 7.} Standardized commissioning \(K\):~
 \(K_{1}\) (Standardized startup
 procedure and attributes support the use of commissioning tools in
 a multi-vendor environment).

 {\it 8.} Robust mesh networking \(F'\):~
      {\it 8.1.} Fault tolerant routing algorithms \(R\):~
      \(R_{1}\) (Response to changes in the network and in the RF
      environment),
        {\it 8.2.} Neighborhood tables \(T\):~
        \(T_{1}\) (Kept by every device).

 {\it 9.} Cluster library support \(L\):~
 \(L_{1}\) (Standardizes application
 behavior across profiles).

 Here two equivalent descriptions of communication protocol
 are considered:
 (a) a structure as a set of protocol components,
 (b) a basic protocol and a set of change (improvement)
 operations.
 In this paper,
 protocol \(S_{3}\) is considered as a basic protocol for forecasting.

 Further, let us consider a direct expert-based forecast
  (version of generation 4)
  ZigBee/IP(6LoWPAN) 2010 (\(\widetilde{S_{4}}\)) as the following:

 {\it 1.} Interference avoidance \(A'\):~
     {\it 1.1.} Startup Procedure of Channel Acquisition \(M\):~
     \(M_{1}\)
     (PAN coordinator selects best available RF channel/Network ID at startup
     time),
     {\it 1.2.} Channel Hopping \(N\):~
     \(N_{1}\) (Ongoing interference detection and adoption of
     a new operating RF channel and/or Network ID).

 {\it 2.} Automated/distributed address management \(B\):~
 \(B_{1}\) (Device
 addresses automatically assigned using a hierarchical,
 distributed scheme),
 \(B_{2}\) (Device addresses automatically assigned using a
 stochastic scheme).

 {\it 3.} Group addressing \(I\):~
 \(I_{1}\) (Devices can be assigned to groups, and
 whole groups can be addressed with a single frame).

  {\it 4.} Centralized data collection \(C''\):~
  {\it 4.1.} Low-overhead data collection by 6LoWPAN Coordinator
  \(G\):~
       \(G_{1}\) (Fully supported),
  {\it 4.2} Low-overhead data collection by other devices \(H\):~
  \(H_{1}\) (Under special circumstances),
  {\it 4.3.} Many-to-one routing \(Q\):~
  \(Q_{1}\) (Whole network discovers the aggregator in one pass),
  and
  {\it 4.4.} 6LoWPAN multicast/broadcast support \(V\):~
   \(V_{1}\) (flooding),
   \(V_{2}\) (unicasting to a PAN coordinator).

 {\it 5.} Network scalability \(D\):~
 \(D_{2}\) (An addressing algorithm that
 relaxes the limits on network size. Networks with hundreds to
 thousands of devices are supported).

 {\it 6.} Message size \(E\):~
 \(E_{3}\) (Large
 messages, up to the buffer capacity of the sending and receiving
 devices using 6LoWPAN Fragmentation and Reassembly).

 {\it 7.} Standardized commissioning \(K\):~
  \(K_{1}\) (Standardized startup procedure
 and attributes support the use of commissioning tools in a
 multi-vendor environment).

 {\it 8.} Robust mesh networking \(F''\):~
 {\it 8.1.} 6LoWPAN Approaches \(U\):~
        \(U_{1}\) (Route-over),
        \(U_{2}\) (Mesh-under).

 {\it 9.} Cluster Library support \(L\):~
 \(L_{1}\) (Standardizes application
 behavior across profiles).

  {\it 10.} Web services support \(W\):~
  \(W_{1}\)
 (condensed HTTP with tokenized XML data).

 Fig. 3, 4, 5, and 6 illustrate the described protocol structures.

\begin{center}
\begin{picture}(52,31)
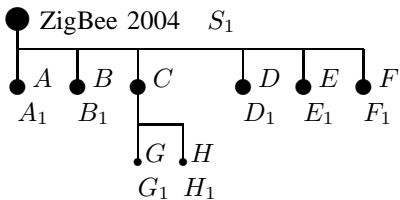
\put(0,0){\makebox(0,0)[bl] {Fig. 3. Structure of generation 1}}

\put(03,27){\makebox(0,0)[bl]{ZigBee 2004 ~ \(S_{1}\)}}

\put(00,29){\circle*{3}}

\put(00,25){\line(0,1){04}}

\put(00,25){\line(1,0){46}}


\put(00,20){\line(0,1){05}}

\put(02,20){\makebox(0,0)[bl]{\(A\)}}

\put(00,20){\circle*{2}}

\put(00,15){\makebox(0,0)[bl]{\(A_{1}\)}}


\put(08,20){\line(0,1){05}}

\put(10,20){\makebox(0,0)[bl]{\(B\)}}

\put(08,20){\circle*{2}}

\put(08,15){\makebox(0,0)[bl]{\(B_{1}\)}}


\put(16,20){\line(0,1){05}}

\put(18,20){\makebox(0,0)[bl]{\(C\)}}

\put(16,20){\circle*{2}}

\put(16,15){\line(0,1){05}}

\put(16,15){\line(1,0){6}}

\put(16,10){\line(0,1){05}} \put(22,10){\line(0,1){05}}

\put(16,10){\circle*{1.2}} \put(22,10){\circle*{1.2}}

\put(17,10){\makebox(0,0)[bl]{\(G\)}}
\put(23,10){\makebox(0,0)[bl]{\(H\)}}

\put(16,05){\makebox(0,0)[bl]{\(G_{1}\)}}

\put(22,05){\makebox(0,0)[bl]{\(H_{1}\)}}


\put(30,20){\line(0,1){05}}

\put(32,20){\makebox(0,0)[bl]{\(D\)}}

\put(30,20){\circle*{2}}

\put(30,15){\makebox(0,0)[bl]{\(D_{1}\)}}


\put(38,20){\line(0,1){05}}

\put(40,20){\makebox(0,0)[bl]{\(E\)}}

\put(38,20){\circle*{2}}

\put(38,15){\makebox(0,0)[bl]{\(E_{1}\)}}


\put(46,20){\line(0,1){05}}

\put(48,20){\makebox(0,0)[bl]{\(F\)}}

\put(46,20){\circle*{2}}

\put(46,15){\makebox(0,0)[bl]{\(F_{1}\)}}

\end{picture}
\end{center}

\begin{center}
\begin{picture}(75,31)
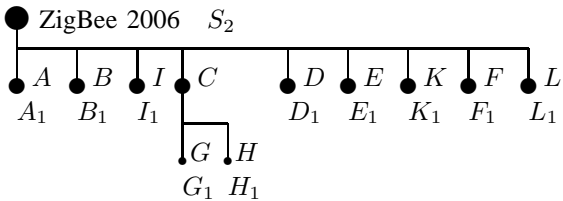
\put(10,0){\makebox(0,0)[bl] {Fig. 4. Structure of generation 2}}

\put(03,27){\makebox(0,0)[bl]{ZigBee 2006 ~ \(S_{2}\)}}

\put(00,29){\circle*{3}}

\put(00,25){\line(0,1){04}}

\put(00,25){\line(1,0){68}}


\put(00,20){\line(0,1){05}}

\put(02,20){\makebox(0,0)[bl]{\(A\)}}

\put(00,20){\circle*{2}}

\put(00,15){\makebox(0,0)[bl]{\(A_{1}\)}}


\put(08,20){\line(0,1){05}}

\put(10,20){\makebox(0,0)[bl]{\(B\)}}

\put(08,20){\circle*{2}}

\put(08,15){\makebox(0,0)[bl]{\(B_{1}\)}}


\put(16,20){\line(0,1){05}}

\put(18,20){\makebox(0,0)[bl]{\(I\)}}

\put(16,20){\circle*{2}}

\put(16,15){\makebox(0,0)[bl]{\(I_{1}\)}}


\put(22,20){\line(0,1){05}}

\put(24,20){\makebox(0,0)[bl]{\(C\)}}

\put(22,20){\circle*{2}}

\put(22,15){\line(0,1){05}}

\put(22,15){\line(1,0){6}}

\put(22,10){\line(0,1){05}} \put(28,10){\line(0,1){05}}

\put(22,10){\circle*{1.2}} \put(28,10){\circle*{1.2}}

\put(23,10){\makebox(0,0)[bl]{\(G\)}}
\put(29,10){\makebox(0,0)[bl]{\(H\)}}

\put(22,05){\makebox(0,0)[bl]{\(G_{1}\)}}

\put(28,05){\makebox(0,0)[bl]{\(H_{1}\)}}


\put(36,20){\line(0,1){05}}

\put(38,20){\makebox(0,0)[bl]{\(D\)}}

\put(36,20){\circle*{2}}

\put(36,15){\makebox(0,0)[bl]{\(D_{1}\)}}


\put(44,20){\line(0,1){05}}

\put(46,20){\makebox(0,0)[bl]{\(E\)}}

\put(44,20){\circle*{2}}

\put(44,15){\makebox(0,0)[bl]{\(E_{1}\)}}


\put(52,20){\line(0,1){05}}

\put(54,20){\makebox(0,0)[bl]{\(K\)}}

\put(52,20){\circle*{2}}

\put(52,15){\makebox(0,0)[bl]{\(K_{1}\)}}


\put(60,20){\line(0,1){05}}

\put(62,20){\makebox(0,0)[bl]{\(F\)}}

\put(60,20){\circle*{2}}

\put(60,15){\makebox(0,0)[bl]{\(F_{1}\)}}


\put(68,20){\line(0,1){05}}

\put(70,20){\makebox(0,0)[bl]{\(L\)}}

\put(68,20){\circle*{2}}

\put(68,15){\makebox(0,0)[bl]{\(L_{1}\)}}

\end{picture}
\end{center}

\begin{center}
\begin{picture}(80,31)
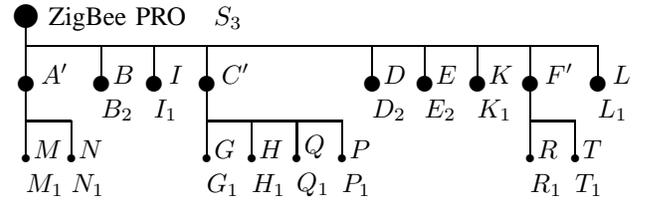
\put(12,0){\makebox(0,0)[bl] {Fig. 5. Structure of generation 3}}

\put(03,27){\makebox(0,0)[bl]{ZigBee PRO ~ \(S_{3}\)}}

\put(00,29){\circle*{3}}

\put(00,25){\line(0,1){04}}

\put(00,25){\line(1,0){76}}


\put(00,20){\line(0,1){05}}

\put(02,20){\makebox(0,0)[bl]{\(A'\)}}

\put(00,20){\circle*{2}}

\put(00,15){\line(0,1){05}}

\put(00,15){\line(1,0){6}}

\put(00,10){\line(0,1){05}} \put(06,10){\line(0,1){05}}

\put(00,10){\circle*{1.2}} \put(06,10){\circle*{1.2}}

\put(01,10){\makebox(0,0)[bl]{\(M\)}}
\put(07,10){\makebox(0,0)[bl]{\(N\)}}

\put(00,05){\makebox(0,0)[bl]{\(M_{1}\)}}

\put(06,05){\makebox(0,0)[bl]{\(N_{1}\)}}


\put(10,20){\line(0,1){05}}

\put(11.5,20){\makebox(0,0)[bl]{\(B\)}}

\put(10,20){\circle*{2}}

\put(10,15){\makebox(0,0)[bl]{\(B_{2}\)}}


\put(17,20){\line(0,1){05}}

\put(19,20){\makebox(0,0)[bl]{\(I\)}}

\put(17,20){\circle*{2}}

\put(17,15){\makebox(0,0)[bl]{\(I_{1}\)}}


\put(24,20){\line(0,1){05}}

\put(26,20){\makebox(0,0)[bl]{\(C'\)}}

\put(24,20){\circle*{2}}

\put(24,15){\line(0,1){05}}

\put(24,15){\line(1,0){18}}

\put(24,10){\line(0,1){05}} \put(30,10){\line(0,1){05}}
\put(36,10){\line(0,1){05}} \put(42,10){\line(0,1){05}}

\put(24,10){\circle*{1.2}} \put(30,10){\circle*{1.2}}
\put(36,10){\circle*{1.2}} \put(42,10){\circle*{1.2}}

\put(25,10){\makebox(0,0)[bl]{\(G\)}}
\put(31,10){\makebox(0,0)[bl]{\(H\)}}
\put(37,10){\makebox(0,0)[bl]{\(Q\)}}
\put(43,10){\makebox(0,0)[bl]{\(P\)}}

\put(24,05){\makebox(0,0)[bl]{\(G_{1}\)}}
\put(30,05){\makebox(0,0)[bl]{\(H_{1}\)}}
\put(36,05){\makebox(0,0)[bl]{\(Q_{1}\)}}
\put(42,05){\makebox(0,0)[bl]{\(P_{1}\)}}


\put(46,20){\line(0,1){05}}

\put(47.5,20){\makebox(0,0)[bl]{\(D\)}}

\put(46,20){\circle*{2}}

\put(46,15){\makebox(0,0)[bl]{\(D_{2}\)}}


\put(53,20){\line(0,1){05}}

\put(54.5,20){\makebox(0,0)[bl]{\(E\)}}

\put(53,20){\circle*{2}}

\put(53,15){\makebox(0,0)[bl]{\(E_{2}\)}}


\put(60,20){\line(0,1){05}}

\put(61.5,20){\makebox(0,0)[bl]{\(K\)}}

\put(60,20){\circle*{2}}

\put(60,15){\makebox(0,0)[bl]{\(K_{1}\)}}


\put(67,20){\line(0,1){05}}

\put(69,20){\makebox(0,0)[bl]{\(F'\)}}

\put(67,20){\circle*{2}}

\put(67,15){\line(0,1){05}}

\put(67,15){\line(1,0){6}}

\put(67,10){\line(0,1){05}}

\put(67,10){\circle*{1.2}}

\put(68,10){\makebox(0,0)[bl]{\(R\)}}
\put(67,05){\makebox(0,0)[bl]{\(R_{1}\)}}

\put(73,10){\line(0,1){05}}

\put(73,10){\circle*{1.2}}

\put(74,10){\makebox(0,0)[bl]{\(T\)}}
\put(73,05){\makebox(0,0)[bl]{\(T_{1}\)}}


\put(76,20){\line(0,1){05}}

\put(78,20){\makebox(0,0)[bl]{\(L\)}}

\put(76,20){\circle*{2}}

\put(76,15){\makebox(0,0)[bl]{\(L_{1}\)}}

\end{picture}
\end{center}

\begin{center}
\begin{picture}(83,36)
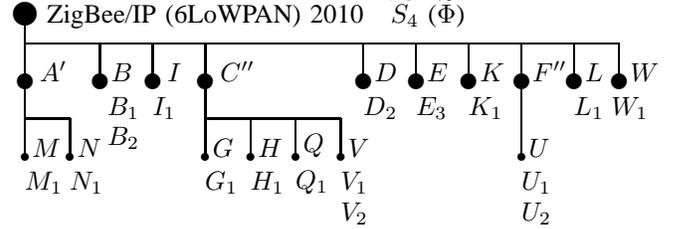
\put(5,00){\makebox(0,0)[bl] {Fig. 6. Structure of direct
expert-based forecast}}

\put(03,32){\makebox(0,0)[bl]{ZigBee/IP (6LoWPAN) 2010
 ~ \(\widetilde{S_{4}}\) (\(\widetilde{\Phi} \))}}

\put(00,34){\circle*{3}}

\put(00,30){\line(0,1){04}}

\put(00,30){\line(1,0){79}}


\put(00,25){\line(0,1){05}}

\put(02,25){\makebox(0,0)[bl]{\(A'\)}}

\put(00,25){\circle*{2}}

\put(00,20){\line(0,1){05}}

\put(00,20){\line(1,0){6}}

\put(00,15){\line(0,1){05}} \put(06,15){\line(0,1){05}}

\put(00,15){\circle*{1.2}} \put(06,15){\circle*{1.2}}

\put(01,15){\makebox(0,0)[bl]{\(M\)}}
\put(07,15){\makebox(0,0)[bl]{\(N\)}}

\put(00,10){\makebox(0,0)[bl]{\(M_{1}\)}}

\put(06,10){\makebox(0,0)[bl]{\(N_{1}\)}}


\put(10,25){\line(0,1){05}}

\put(11.5,25){\makebox(0,0)[bl]{\(B\)}}

\put(10,25){\circle*{2}}

\put(11,20){\makebox(0,0)[bl]{\(B_{1}\)}}
\put(11,16){\makebox(0,0)[bl]{\(B_{2}\)}}


\put(17,25){\line(0,1){05}}

\put(19,25){\makebox(0,0)[bl]{\(I\)}}

\put(17,25){\circle*{2}}

\put(17,20){\makebox(0,0)[bl]{\(I_{1}\)}}


\put(24,25){\line(0,1){05}}

\put(26,25){\makebox(0,0)[bl]{\(C''\)}}

\put(24,25){\circle*{2}}

\put(24,20){\line(0,1){05}}

\put(24,20){\line(1,0){18}}

\put(24,15){\line(0,1){05}} \put(30,15){\line(0,1){05}}
\put(36,15){\line(0,1){05}} \put(42,15){\line(0,1){05}}

\put(24,15){\circle*{1.2}} \put(30,15){\circle*{1.2}}
\put(36,15){\circle*{1.2}} \put(42,15){\circle*{1.2}}

\put(25,15){\makebox(0,0)[bl]{\(G\)}}
\put(31,15){\makebox(0,0)[bl]{\(H\)}}
\put(37,15){\makebox(0,0)[bl]{\(Q\)}}
\put(43,15){\makebox(0,0)[bl]{\(V\)}}

\put(24,10){\makebox(0,0)[bl]{\(G_{1}\)}}
\put(30,10){\makebox(0,0)[bl]{\(H_{1}\)}}
\put(36,10){\makebox(0,0)[bl]{\(Q_{1}\)}}
\put(42,10){\makebox(0,0)[bl]{\(V_{1}\)}}
\put(42,06){\makebox(0,0)[bl]{\(V_{2}\)}}


\put(45,25){\line(0,1){05}}

\put(46.5,25){\makebox(0,0)[bl]{\(D\)}}

\put(45,25){\circle*{2}}

\put(45,20){\makebox(0,0)[bl]{\(D_{2}\)}}


\put(52,25){\line(0,1){05}}

\put(53.5,25){\makebox(0,0)[bl]{\(E\)}}

\put(52,25){\circle*{2}}

\put(52,20){\makebox(0,0)[bl]{\(E_{3}\)}}


\put(59,25){\line(0,1){05}}

\put(60.5,25){\makebox(0,0)[bl]{\(K\)}}

\put(59,25){\circle*{2}}

\put(59,20){\makebox(0,0)[bl]{\(K_{1}\)}}


\put(66,25){\line(0,1){05}}

\put(67.5,25){\makebox(0,0)[bl]{\(F''\)}}

\put(66,25){\circle*{2}}

\put(66,20){\line(0,1){05}}


\put(66,15){\line(0,1){05}}

\put(66,15){\circle*{1.2}}

\put(67,15){\makebox(0,0)[bl]{\(U\)}}
\put(66,10){\makebox(0,0)[bl]{\(U_{1}\)}}
\put(66,06){\makebox(0,0)[bl]{\(U_{2}\)}}


\put(73,25){\line(0,1){05}}

\put(74.5,25){\makebox(0,0)[bl]{\(L\)}}

\put(73,25){\circle*{2}}

\put(73,20){\makebox(0,0)[bl]{\(L_{1}\)}}


\put(79,25){\line(0,1){05}}

\put(80.5,25){\makebox(0,0)[bl]{\(W\)}}

\put(79,25){\circle*{2}}

\put(78,20){\makebox(0,0)[bl]{\(W_{1}\)}}

\end{picture}
\end{center}

\section{Change Operations}

 Table 1 integrates changes in protocol generations.

\begin{center}
\begin{picture}(80,116)

\put(08,112){\makebox(0,0)[bl]{Table 1. Changes in protocol
generations
}}


\put(00,0){\line(1,0){80}} \put(00,98){\line(1,0){80}}
\put(00,110){\line(1,0){80}}

\put(00,0){\line(0,1){110}} \put(21,98){\line(0,1){12}}
\put(68,98){\line(0,1){12}} \put(80,0){\line(0,1){110}}


\put(01,106){\makebox(0,0)[bl]{Kind of}}
\put(01,102.5){\makebox(0,0)[bl]{change}}

\put(22,106){\makebox(0,0)[bl]{Change}}


\put(68.5,106){\makebox(0,0)[bl]{Opera-}}
\put(68.5,103.5){\makebox(0,0)[bl]{tion}}
\put(68.5,100){\makebox(0,0)[bl]{type}}


\put(01,93){\makebox(0,0)[bl]{\(S_{1} \Longrightarrow S_{2}\)}}





\put(01,89){\makebox(0,0)[bl]{2.Subsystem }}

\put(04,86){\makebox(0,0)[bl]{addition}}


\put(22,89){\makebox(0,0)[bl]{(a) \(I\), \(I_{1}\)}}

\put(22,85){\makebox(0,0)[bl]{(b) \(K\), \(K_{1}\)}}

\put(22,81){\makebox(0,0)[bl]{(c) \(L\), \(L_{1}\)}}

\put(72,89){\makebox(0,0)[bl]{\(O_{7}\)}}
\put(72,85){\makebox(0,0)[bl]{\(O_{7}\)}}
\put(72,81){\makebox(0,0)[bl]{\(O_{7}\)}}


\put(00,79){\line(1,0){80}}

\put(01,75){\makebox(0,0)[bl]{\(S_{2} \Longrightarrow S_{3}\)}}

\put(01,70){\makebox(0,0)[bl]{1.Element}}
\put(04,66.5){\makebox(0,0)[bl]{change}}


\put(22,70){\makebox(0,0)[bl]{(a) \(B_{1} \rightarrow B_{2}\)}}
\put(22,66){\makebox(0,0)[bl]{(b) \(D_{1} \rightarrow D_{2}\)}}
\put(22,62){\makebox(0,0)[bl]{(c) \(E_{1} \rightarrow E_{2}\)}}

\put(72,70){\makebox(0,0)[bl]{\(O_{1}\)}}
\put(72,66){\makebox(0,0)[bl]{\(O_{1}\)}}
\put(72,62){\makebox(0,0)[bl]{\(O_{1}\)}}


\put(01,58){\makebox(0,0)[bl]{2.Subsystem}}
\put(04,55.4){\makebox(0,0)[bl]{extension}}


\put(22,58){\makebox(0,0)[bl]{(a) \(C \rightarrow C'\):}}

\put(45,58){\makebox(0,0)[bl]{(i) \(Q\), \(Q_{1}\)}}

\put(45,54){\makebox(0,0)[bl]{(ii) \(P\), \(P_{1}\)}}

\put(72,58){\makebox(0,0)[bl]{\(O_{5}\)}}
\put(72,54){\makebox(0,0)[bl]{\(O_{5}\)}}


\put(22,50){\makebox(0,0)[bl]{(b) \(A \rightarrow A'\):}}

\put(45,50){\makebox(0,0)[bl]{(i) \(M\), \(M_{1}\)}}

\put(45,46){\makebox(0,0)[bl]{(ii) \(N\), \(N_{1}\)}}

\put(72,50){\makebox(0,0)[bl]{\(O_{5}\)}}
\put(72,46){\makebox(0,0)[bl]{\(O_{5}\)}}


\put(22,42){\makebox(0,0)[bl]{(c) \(F \rightarrow F'\):}}

\put(45,42){\makebox(0,0)[bl]{(i) \(R\), \(R_{1}\)}}

\put(45,38){\makebox(0,0)[bl]{(ii) \(T\), \(T_{1}\)}}

\put(72,42){\makebox(0,0)[bl]{\(O_{5}\)}}
\put(72,38){\makebox(0,0)[bl]{\(O_{5}\)}}


\put(00,36){\line(1,0){80}}

\put(01,31){\makebox(0,0)[bl]{\(S_{3} \Longrightarrow
\widetilde{S_{4}}\)}}

\put(01,27){\makebox(0,0)[bl]{1.Element}}
\put(04,24){\makebox(0,0)[bl]{addition}}


\put(22,27){\makebox(0,0)[bl]{\(B_{1}\)}}

\put(72,27){\makebox(0,0)[bl]{\(O_{3}\)}}


\put(01,20){\makebox(0,0)[bl]{2.Element}}
\put(04,17){\makebox(0,0)[bl]{change}}

\put(22,20){\makebox(0,0)[bl]{\(E_{2} \rightarrow E_{3}\)}}

\put(72,20){\makebox(0,0)[bl]{\(O_{1}\)}}


\put(01,13){\makebox(0,0)[bl]{3.Subsystem}}
\put(04,10){\makebox(0,0)[bl]{addition}}

\put(22,13){\makebox(0,0)[bl]{\(W\), \(W_{1}\)}}

\put(72,13){\makebox(0,0)[bl]{\(O_{7}\)}}


\put(01,06){\makebox(0,0)[bl]{4.Subsystem}}
\put(04,03){\makebox(0,0)[bl]{change}}

\put(22,06){\makebox(0,0)[bl]{(a) \(C' \Rightarrow C'':\) }}

\put(45,06){\makebox(0,0)[bl]{\(P \rightarrow V,  ~ V_{1}, V_{2}\)
}}

\put(22,02){\makebox(0,0)[bl]{(b) \(F' \Rightarrow F'':\)}}
\put(45,02){\makebox(0,0)[bl]{\(U,  ~ U_{1},  U_{2}\)}}

\put(72,06){\makebox(0,0)[bl]{\(O_{5}\)}}
\put(72,02){\makebox(0,0)[bl]{\(O_{5}\)}}

\end{picture}
\end{center}

 Now it is necessary to generate a basic set of possible
 change/improvement operations.
 This process is based on the following:
 (a) obtained protocol changes (Table 1),
 (b) additional expert judgement.
 Thus, the resultant set of the possible operations is;

 {\it 1.} \( \Phi_{1}\):~ \(I_{1}\).
 Introduction of groups allows to transmit a single frame to
 all devices assigned to a group. It has its positive impact on
 scalability and reliability of the network.
 The cost and the implementation time are negatively related
 to device association list introduction.
 The maintenance time will remain the same because of
 the group maintenance tasks time reduction compensation by
 an additional time required for handling the device association lists.

 {\it 2.} \( \Phi_{2}\):~ \(K_{1}\).
 Standardized commissioning decreases maintenance efforts an cost;
 increases scalability and reliability.

 {\it 3.} \( \Phi_{3}\):~ \(L_{1}\).
 ZigBee Cluster Library standardizes application behavior
 resulting in better reliability and lower maintenance efforts.

 {\it 4.} \( \Phi_{4}\):~ \(B_{1} \rightarrow B_{2}\).
 Stochastic device address management does not require
 the knowledge of network hierarchy resulting in improved
 reliability and mobility.
 While increasing the risk of collision it allows to assign more
 addresses.

 {\it 5.} \( \Phi_{5}\):~ \(D_{1} \rightarrow D_{2}\).
 Addressing algorithm limits relaxation drastically
 increases scalability.

 {\it 6.} \( \Phi_{6}\):~ \(E_{1} \rightarrow E_{2}\).
  Message size enlargement will reduce the time
 necessary for WSN application development and maintenance while
 improving the reliability.

 {\it 7.} \( \Phi_{7}\):~ \(Q_{1}\).
 Many-to-one routing reduces application implementation time
 but can cause aggregator buffer overflow.

 {\it 8.} \( \Phi_{8}\):~ \(P_{1}\).
 Source routing also reduces application implementation time
 but improving the reliability.
 Source routing allows easier troubleshooting, improved trace-route,
 and enables a node to discover all the possible routes to a host.
 It also allows a source to directly manage network performance by
 forcing packets to travel over one path to prevent congestion on another.

 {\it 9.} \( \Phi_{9}\):~ \(M_{1}\).
 Startup channel acquisition procedure is an interference
 avoidance mechanism requiring an additional resources but
 improving reliability, mobility, scalability and optimizing
 maintenance efforts.

 {\it 10.} \( \Phi_{10}\):~ \(N_{1}\).
 Channel hopping requires more resources but brings an
 additional mobility, reliability while reducing maintenance efforts.

  {\it 11.} \( \Phi_{11}\):~ \(R_{1}\).
 Fault tolerant routing algorithms aims at reliability
  and
  mobility.

 {\it 12.} \( \Phi_{12}\):~ \(T_{1}\).
 Neighborhood tables need memory but positively influence on
 scalability, reliability and mobility.

 {\it 13.} \( \Phi_{13}\):~ \(B_{1}\).
 The combination of both address distribution schemes
 increase mobility of nodes while keeping maintenance costs at
 acceptable level.

 {\it 14.} \( \Phi_{14}\):~ \(E_{2} \rightarrow E_{3}\).
 Providing large message sizes by 6LoWPAN
 fragmentation and reassembly mechanisms significantly improves
 scalability through heterogeneous WSNs.

 {\it 15.} \( \Phi_{15}\):~ \(W_{1}\).
 Porting HTTP to WSN level is a significant step towards
 ubiquituos user-friendly data propagation.

 {\it 16.} \( \Phi_{16}\):~
 \(P \rightarrow V\), \(V_{1}\), \(V_{2}\).
 6LoWPAN multicast/broadcast support will reduce
 the development time needed in heterogeneous WSNs, increase
 overall system reliability and usability.

 {\it 17.} \( \Phi_{17}\):~
 \(U_{1}\), \(U_{2}\). 6LoWPAN mesh networking approaches are necessary to
 provide an interoperable platform for heterogeneous WSNs
 that would lead to better scalability.

 Here the following attributes (criteria) for an assessment of the
 operations are used:
 (1)  cost
 \(\Upsilon_{1}\);
 (2) required time for implementation \(\Upsilon_{2}\);
 (3) performance \(\Upsilon_{3}\);
 (4) decreasing a cost of  maintenance \(\Upsilon_{4}\);
 (5) scalability  \(\Upsilon_{5}\);
 (6) reliability \(\Upsilon_{6}\);
 (7) mobility \(\Upsilon_{7}\); and
 (8) usability value \(\Upsilon_{8}\).
 An ordinal scale [1,5] is used for each criterion:
 \(1\) corresponds to ``strong negative effect'',
  \(2\) corresponds to ``negative effect'',
   \(3\) corresponds to ``no changes'',
 \(4\) corresponds to ``positive effect'', and
 \(5\) corresponds to ``strong positive effect''.

 Table 2 contains improvement operations
  ~\(\Phi_{1}\), ..., \(\Phi_{i}\), ..., \(\Phi_{17}\)~
 and their estimates upon criteria.

\begin{center}
\begin{picture}(80,87)

\put(08.5,83){\makebox(0,0)[bl] {Table 2. Estimates on criteria,
priorities}}


\put(0,0){\line(1,0){80}} \put(0,71){\line(1,0){80}}
\put(0,81){\line(1,0){80}}

\put(0,0){\line(0,1){81}} \put(21,0){\line(0,1){81}}
\put(69,0){\line(0,1){81}} \put(80,0){\line(0,1){81}}

\put(27,71){\line(0,1){10}} \put(33,71){\line(0,1){10}}
\put(39,71){\line(0,1){10}}

\put(45,71){\line(0,1){10}} \put(51,71){\line(0,1){10}}
\put(57,71){\line(0,1){10}} \put(63,71){\line(0,1){10}}

\put(0.5,76.5){\makebox(0,0)[bl]{Improvement}}
\put(0.5,72.5){\makebox(0,0)[bl]{operation}}

\put(22,76){\makebox(0,0)[bl]{\(\Upsilon_{1}\)}}
\put(28,76){\makebox(0,0)[bl]{\(\Upsilon_{2}\)}}
\put(34,76){\makebox(0,0)[bl]{\(\Upsilon_{3}\)}}
\put(40,76){\makebox(0,0)[bl]{\(\Upsilon_{4}\)}}

\put(46,76){\makebox(0,0)[bl]{\(\Upsilon_{5}\)}}
\put(52,76){\makebox(0,0)[bl]{\(\Upsilon_{6}\)}}
\put(58,76){\makebox(0,0)[bl]{\(\Upsilon_{7}\)}}
\put(64,76){\makebox(0,0)[bl]{\(\Upsilon_{8}\)}}

\put(69.5,77){\makebox(0,0)[bl]{Priori-}}
\put(69.5,73){\makebox(0,0)[bl]{ties  \(r_{i}\)}}


\put(8,66){\makebox(0,0)[bl]{\(\Phi_{1}\)}}
\put(8,62){\makebox(0,0)[bl]{\(\Phi_{2}\)}}
\put(8,58){\makebox(0,0)[bl]{\(\Phi_{3}\)}}
\put(8,54){\makebox(0,0)[bl]{\(\Phi_{4}\)}}
\put(8,50){\makebox(0,0)[bl]{\(\Phi_{5}\)}}
\put(8,46){\makebox(0,0)[bl]{\(\Phi_{6}\)}}
\put(8,42){\makebox(0,0)[bl]{\(\Phi_{7}\)}}
\put(8,38){\makebox(0,0)[bl]{\(\Phi_{8}\)}}
\put(8,34){\makebox(0,0)[bl]{\(\Phi_{9}\)}}
\put(8,30){\makebox(0,0)[bl]{\(\Phi_{10}\)}}
\put(8,26){\makebox(0,0)[bl]{\(\Phi_{11}\)}}
\put(8,22){\makebox(0,0)[bl]{\(\Phi_{12}\)}}
\put(8,18){\makebox(0,0)[bl]{\(\Phi_{13}\)}}
\put(8,14){\makebox(0,0)[bl]{\(\Phi_{14}\)}}
\put(8,10){\makebox(0,0)[bl]{\(\Phi_{15}\)}}
\put(8,06){\makebox(0,0)[bl]{\(\Phi_{16}\)}}
\put(8,02){\makebox(0,0)[bl]{\(\Phi_{17}\)}}

\put(23,66){\makebox(0,0)[bl]{\(3\)}}
\put(29,66){\makebox(0,0)[bl]{\(3\)}}

\put(35,66){\makebox(0,0)[bl]{\(3\)}}

\put(41,66){\makebox(0,0)[bl]{\(3\)}}
\put(47,66){\makebox(0,0)[bl]{\(4\)}}

\put(53,66){\makebox(0,0)[bl]{\(4\)}}
\put(59,66){\makebox(0,0)[bl]{\(3\)}}
\put(65,66){\makebox(0,0)[bl]{\(4\)}}

\put(73.5,66){\makebox(0,0)[bl]{\(1\)}}

\put(23,62){\makebox(0,0)[bl]{\(4\)}}
\put(29,62){\makebox(0,0)[bl]{\(2\)}}

\put(35,62){\makebox(0,0)[bl]{\(3\)}}

\put(41,62){\makebox(0,0)[bl]{\(4\)}}
\put(47,62){\makebox(0,0)[bl]{\(4\)}}

\put(53,62){\makebox(0,0)[bl]{\(4\)}}
\put(59,62){\makebox(0,0)[bl]{\(3\)}}
\put(65,62){\makebox(0,0)[bl]{\(4\)}}

\put(73.5,62){\makebox(0,0)[bl]{\(2\)}}

\put(23,58){\makebox(0,0)[bl]{\(3\)}}
\put(29,58){\makebox(0,0)[bl]{\(4\)}}

\put(35,58){\makebox(0,0)[bl]{\(3\)}}

\put(41,58){\makebox(0,0)[bl]{\(4\)}}
\put(47,58){\makebox(0,0)[bl]{\(3\)}}

\put(53,58){\makebox(0,0)[bl]{\(5\)}}
\put(59,58){\makebox(0,0)[bl]{\(3\)}}
\put(65,58){\makebox(0,0)[bl]{\(4\)}}

\put(73.5,58){\makebox(0,0)[bl]{\(1\)}}

\put(23,54){\makebox(0,0)[bl]{\(4\)}}
\put(29,54){\makebox(0,0)[bl]{\(4\)}}

\put(35,54){\makebox(0,0)[bl]{\(2\)}}

\put(41,54){\makebox(0,0)[bl]{\(3\)}}
\put(47,54){\makebox(0,0)[bl]{\(4\)}}

\put(53,54){\makebox(0,0)[bl]{\(4\)}}
\put(59,54){\makebox(0,0)[bl]{\(5\)}}
\put(65,54){\makebox(0,0)[bl]{\(3\)}}

\put(73.5,54){\makebox(0,0)[bl]{\(2\)}}

\put(23,50){\makebox(0,0)[bl]{\(3\)}}
\put(29,50){\makebox(0,0)[bl]{\(3\)}}

\put(35,50){\makebox(0,0)[bl]{\(3\)}}

\put(41,50){\makebox(0,0)[bl]{\(3\)}}
\put(47,50){\makebox(0,0)[bl]{\(5\)}}

\put(53,50){\makebox(0,0)[bl]{\(3\)}}
\put(59,50){\makebox(0,0)[bl]{\(3\)}}
\put(65,50){\makebox(0,0)[bl]{\(4\)}}

\put(73.5,50){\makebox(0,0)[bl]{\(1\)}}

\put(23,46){\makebox(0,0)[bl]{\(3\)}}
\put(29,46){\makebox(0,0)[bl]{\(4\)}}

\put(35,46){\makebox(0,0)[bl]{\(3\)}}

\put(41,46){\makebox(0,0)[bl]{\(4\)}}
\put(47,46){\makebox(0,0)[bl]{\(3\)}}

\put(53,46){\makebox(0,0)[bl]{\(4\)}}
\put(59,46){\makebox(0,0)[bl]{\(3\)}}
\put(65,46){\makebox(0,0)[bl]{\(3\)}}

\put(73.5,46){\makebox(0,0)[bl]{\(3\)}}

\put(23,42){\makebox(0,0)[bl]{\(3\)}}
\put(29,42){\makebox(0,0)[bl]{\(4\)}}

\put(35,42){\makebox(0,0)[bl]{\(3\)}}

\put(41,42){\makebox(0,0)[bl]{\(3\)}}
\put(47,42){\makebox(0,0)[bl]{\(3\)}}

\put(53,42){\makebox(0,0)[bl]{\(2\)}}
\put(59,42){\makebox(0,0)[bl]{\(3\)}}
\put(65,42){\makebox(0,0)[bl]{\(3\)}}

\put(73.5,42){\makebox(0,0)[bl]{\(4\)}}

\put(23,38){\makebox(0,0)[bl]{\(3\)}}
\put(29,38){\makebox(0,0)[bl]{\(4\)}}

\put(35,38){\makebox(0,0)[bl]{\(4\)}}

\put(41,38){\makebox(0,0)[bl]{\(4\)}}
\put(47,38){\makebox(0,0)[bl]{\(3\)}}

\put(53,38){\makebox(0,0)[bl]{\(4\)}}
\put(59,38){\makebox(0,0)[bl]{\(3\)}}
\put(65,38){\makebox(0,0)[bl]{\(3\)}}

\put(73.5,38){\makebox(0,0)[bl]{\(2\)}}

\put(23,34){\makebox(0,0)[bl]{\(2\)}}
\put(29,34){\makebox(0,0)[bl]{\(3\)}}

\put(35,34){\makebox(0,0)[bl]{\(4\)}}

\put(41,34){\makebox(0,0)[bl]{\(4\)}}
\put(47,34){\makebox(0,0)[bl]{\(4\)}}

\put(53,34){\makebox(0,0)[bl]{\(4\)}}
\put(59,34){\makebox(0,0)[bl]{\(4\)}}
\put(65,34){\makebox(0,0)[bl]{\(3\)}}

\put(73.5,34){\makebox(0,0)[bl]{\(1\)}}

\put(23,30){\makebox(0,0)[bl]{\(2\)}}
\put(29,30){\makebox(0,0)[bl]{\(3\)}}

\put(35,30){\makebox(0,0)[bl]{\(4\)}}

\put(41,30){\makebox(0,0)[bl]{\(4\)}}
\put(47,30){\makebox(0,0)[bl]{\(3\)}}

\put(53,30){\makebox(0,0)[bl]{\(4\)}}
\put(59,30){\makebox(0,0)[bl]{\(4\)}}
\put(65,30){\makebox(0,0)[bl]{\(3\)}}

\put(73.5,30){\makebox(0,0)[bl]{\(1\)}}

\put(23,26){\makebox(0,0)[bl]{\(3\)}}
\put(29,26){\makebox(0,0)[bl]{\(3\)}}

\put(35,26){\makebox(0,0)[bl]{\(3\)}}

\put(41,26){\makebox(0,0)[bl]{\(3\)}}
\put(47,26){\makebox(0,0)[bl]{\(3\)}}

\put(53,26){\makebox(0,0)[bl]{\(4\)}}
\put(59,26){\makebox(0,0)[bl]{\(4\)}}
\put(65,26){\makebox(0,0)[bl]{\(3\)}}

\put(73.5,26){\makebox(0,0)[bl]{\(3\)}}

\put(23,22){\makebox(0,0)[bl]{\(2\)}}
\put(29,22){\makebox(0,0)[bl]{\(3\)}}

\put(35,22){\makebox(0,0)[bl]{\(4\)}}

\put(41,22){\makebox(0,0)[bl]{\(3\)}}
\put(47,22){\makebox(0,0)[bl]{\(4\)}}

\put(53,22){\makebox(0,0)[bl]{\(4\)}}
\put(59,22){\makebox(0,0)[bl]{\(4\)}}
\put(65,22){\makebox(0,0)[bl]{\(3\)}}

\put(73.5,22){\makebox(0,0)[bl]{\(1\)}}

\put(23,18){\makebox(0,0)[bl]{\(3\)}}
\put(29,18){\makebox(0,0)[bl]{\(3\)}}

\put(35,18){\makebox(0,0)[bl]{\(3\)}}

\put(41,18){\makebox(0,0)[bl]{\(4\)}}
\put(47,18){\makebox(0,0)[bl]{\(3\)}}

\put(53,18){\makebox(0,0)[bl]{\(3\)}}
\put(59,18){\makebox(0,0)[bl]{\(4\)}}
\put(65,18){\makebox(0,0)[bl]{\(3\)}}

\put(73.5,18){\makebox(0,0)[bl]{\(3\)}}

\put(23,14){\makebox(0,0)[bl]{\(3\)}}
\put(29,14){\makebox(0,0)[bl]{\(3\)}}

\put(35,14){\makebox(0,0)[bl]{\(3\)}}

\put(41,14){\makebox(0,0)[bl]{\(3\)}}
\put(47,14){\makebox(0,0)[bl]{\(5\)}}

\put(53,14){\makebox(0,0)[bl]{\(3\)}}
\put(59,14){\makebox(0,0)[bl]{\(3\)}}
\put(65,14){\makebox(0,0)[bl]{\(3\)}}

\put(73.5,14){\makebox(0,0)[bl]{\(2\)}}

\put(23,10){\makebox(0,0)[bl]{\(3\)}}
\put(29,10){\makebox(0,0)[bl]{\(3\)}}

\put(35,10){\makebox(0,0)[bl]{\(2\)}}

\put(41,10){\makebox(0,0)[bl]{\(3\)}}
\put(47,10){\makebox(0,0)[bl]{\(4\)}}

\put(53,10){\makebox(0,0)[bl]{\(3\)}}
\put(59,10){\makebox(0,0)[bl]{\(3\)}}
\put(65,10){\makebox(0,0)[bl]{\(5\)}}

\put(73.5,10){\makebox(0,0)[bl]{\(2\)}}

\put(23,6){\makebox(0,0)[bl]{\(3\)}}
\put(29,6){\makebox(0,0)[bl]{\(4\)}}

\put(35,6){\makebox(0,0)[bl]{\(3\)}}

\put(41,6){\makebox(0,0)[bl]{\(3\)}}
\put(47,6){\makebox(0,0)[bl]{\(3\)}}

\put(53,6){\makebox(0,0)[bl]{\(4\)}}
\put(59,6){\makebox(0,0)[bl]{\(3\)}}
\put(65,6){\makebox(0,0)[bl]{\(4\)}}

\put(73.5,6){\makebox(0,0)[bl]{\(3\)}}

\put(23,2){\makebox(0,0)[bl]{\(3\)}}
\put(29,2){\makebox(0,0)[bl]{\(3\)}}

\put(35,2){\makebox(0,0)[bl]{\(4\)}}

\put(41,2){\makebox(0,0)[bl]{\(3\)}}
\put(47,2){\makebox(0,0)[bl]{\(4\)}}

\put(53,2){\makebox(0,0)[bl]{\(3\)}}
\put(59,2){\makebox(0,0)[bl]{\(3\)}}
\put(65,2){\makebox(0,0)[bl]{\(3\)}}

\put(73.5,2){\makebox(0,0)[bl]{\(3\)}}

\end{picture}
\end{center}

 In addition, it is reasonable to
 consider some
 types of binary relations over the improvement operations
 (e.g., {\it equivalence}, {\it complementarity}, {\it precedence}).

\section{Computation of Forecasts}

 Results of multicriteria ranking for operations
 \(\{\Phi_{1},...,\Phi_{17} \}\)
  are
 presented in Table 2 (Fig. 7)
 (an outranking technique was used;
 \(1\) corresponds to the best level).
 Priorities of operations can be used
 as a ``profit''
 (here:  ~\(c_{i} =4-r_{i} \)).

\begin{center}
\begin{picture}(60,43)
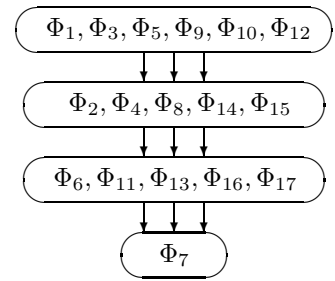
\put(0,0){\makebox(0,0)[bl]{Fig. 7. Results of multicriteria
ranking }}


\put(30,38){\oval(42,6)}

\put(13,36.5){\makebox(0,8)[bl]{\(\Phi_{1},\Phi_{3},\Phi_{5},\Phi_{9},\Phi_{10},\Phi_{12}
\)}}

\put(26,35){\vector(0,-1){4}} \put(30,35){\vector(0,-1){4}}
\put(34,35){\vector(0,-1){4}}


\put(30,28){\oval(40,6)}

\put(16,26.5){\makebox(0,8)[bl]{\(
\Phi_{2},\Phi_{4},\Phi_{8},\Phi_{14},\Phi_{15} \)}}

\put(26,25){\vector(0,-1){4}} \put(30,25){\vector(0,-1){4}}
\put(34,25){\vector(0,-1){4}}


\put(30,18){\oval(40,6)}
\put(14,16.5){\makebox(0,8)[bl]{\(\Phi_{6},\Phi_{11},\Phi_{13},\Phi_{16},\Phi_{17}\)}}

\put(26,15){\vector(0,-1){4}} \put(30,15){\vector(0,-1){4}}
\put(34,15){\vector(0,-1){4}}


\put(30,08){\oval(14,6)}
\put(28,06.5){\makebox(0,8)[bl]{\(\Phi_{7} \)}}

\end{picture}
\end{center}


 Further, let us  consider the usage of knapsack problem:~
 \[\max\sum_{i=1}^{17} c_{i} x_{i}
 ~~s.t. \sum_{i=1}^{17}~ a_{i} x_{i} \leq b,
 ~~x_{i} \in \{ 0, 1\}, ~i=\overline{1,17}.\]
%
%
 For an assessment of resource requirements (i.e., \(a_{i}\))
  the following estimates (additional expert judgment)
 are used:
 \(\{ 2, 3, 4, 1, 1, 2, 2, 3, 2, 4, 3, 3, 3, 3, 3, 2, 4  \}\).
 In this case,
 independence of improvement operations is assumed.
 Thus, the following solution (forecast) is obtained
  (total cost constraint \( b = 16  \)):~
 ~\( \widehat{\Phi}  =
  \{ \Phi_{1},\Phi_{2},\Phi_{4},\Phi_{5},\Phi_{6},\Phi_{7},\Phi_{8},\Phi_{9}
  \}\)
 (a simple greedy algorithm was used).


 Results of clustering for operations
 \(\{\Phi_{1},...,\Phi_{17} \}\)
 (9 steps of an agglomerative algorithm) are the following:
%
 (i) cluster 1:~   \(\Omega^{1} = \{ \Phi_{1},\Phi_{3},\Phi_{6},\Phi_{8},\Phi_{16} \}\),
 (ii) cluster 2:~  \(\Omega^{2} = \{ \Phi_{2} \}\),
 (iii) cluster 3:~ \(\Omega^{3} = \{ \Phi_{4} \}\),
 (iv) cluster 4:~  \(\Omega^{4} = \{ \Phi_{5},\Phi_{14},\Phi_{17}
 \}\),
 (v) cluster 5:~   \(\Omega^{5} = \{ \Phi_{7} \}\),
 (vi) cluster 6:~
 \(\Omega^{6} = \{ \Phi_{9}, \Phi_{10}, \Phi_{12} \}\),
 (vi) cluster 7:~
 \(\Omega^{7} = \{ \Phi_{11}, \Phi_{13} \}\), and
 (vii) cluster 8:~ \(\Omega^{8} = \{ \Phi_{15} \}\).
%
%
 Now it is possible to examine
 multiple choice knapsack problem.
 It is assumed that operations which belong to the same cluster
 are very close (about equivalent) and
 the only one operation from each cluster is selected (if it is possible by resource
 constraint).
 The problem formulation is:
 \[\max\sum_{i=1}^{8} \sum_{j=1}^{q_{i}} c_{ij} x_{ij},
%
 ~~s.t.~\sum_{i=1}^{8} \sum_{j=1}^{q_{i}} a_{ij} x_{ij} \leq
 b,\]
 \[\sum_{j=1}^{q_{i}} x_{ij}=1 ~~i=\overline{1,8},
 ~~x_{ij} \in \{0,1\}.\]
 Priorities and resource requirements
 are examined as in knapsack problem.
 A resultant solution (forecast) is the following
 (total cost constraint \( b = 17  \)):
 ~\( \overline{\Phi}  =
  \{ \Phi_{2},\Phi_{4},\Phi_{5},\Phi_{6},\Phi_{7},\Phi_{9},\Phi_{11},\Phi_{15}
  \}\)
  (a simple greedy algorithm was used).

 Fig. 8 and Fig. 9 depict structures which illustrate solutions
 (i.e., corresponding groups of operations)
 for
 ~\( \widehat{\Phi^{2}}\) and
 ~\( \overline{\Phi^{2}} \).

\begin{center}
\begin{picture}(76,31)
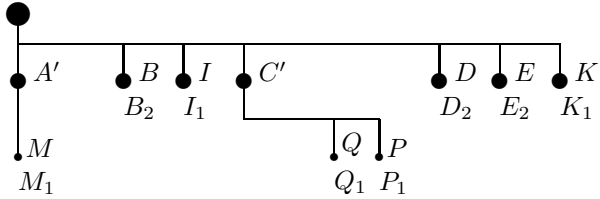
\put(05,0){\makebox(0,0)[bl] {Fig. 8. Forecast ~\(
\widehat{\Phi}\)~ (knapsack problem)}}

\put(03,27){\makebox(0,0)[bl]{
}}

\put(00,29){\circle*{3}}

\put(00,25){\line(0,1){04}}


\put(00,25){\line(1,0){72}}


\put(00,20){\line(0,1){05}}

\put(02,20){\makebox(0,0)[bl]{\(A'\)}}

\put(00,20){\circle*{2}}

\put(00,15){\line(0,1){05}}


\put(00,10){\line(0,1){05}}


\put(00,10){\circle*{1.2}}


\put(01,10){\makebox(0,0)[bl]{\(M\)}}

\put(00,05){\makebox(0,0)[bl]{\(M_{1}\)}}



\put(14,20){\line(0,1){05}}

\put(16,20){\makebox(0,0)[bl]{\(B\)}}

\put(14,20){\circle*{2}}

\put(14,15){\makebox(0,0)[bl]{\(B_{2}\)}}


\put(22,20){\line(0,1){05}}

\put(24,20){\makebox(0,0)[bl]{\(I\)}}

\put(22,20){\circle*{2}}

\put(22,15){\makebox(0,0)[bl]{\(I_{1}\)}}


\put(30,20){\line(0,1){05}}


\put(32,20){\makebox(0,0)[bl]{\(C'\)}}

\put(30,20){\circle*{2}}

\put(30,15){\line(0,1){05}}

\put(30,15){\line(1,0){18}}

\put(30,15){\line(1,0){12}}


\put(42,10){\line(0,1){05}}

\put(48,10){\line(0,1){05}}


\put(42,10){\circle*{1.2}}

\put(48,10){\circle*{1.2}}

\put(43,10){\makebox(0,0)[bl]{\(Q\)}}
\put(49,10){\makebox(0,0)[bl]{\(P\)}}


\put(42,05){\makebox(0,0)[bl]{\(Q_{1}\)}}

\put(48,05){\makebox(0,0)[bl]{\(P_{1}\)}}


\put(56,20){\line(0,1){05}}

\put(58,20){\makebox(0,0)[bl]{\(D\)}}

\put(56,20){\circle*{2}}

\put(56,15){\makebox(0,0)[bl]{\(D_{2}\)}}


\put(64,20){\line(0,1){05}}

\put(66,20){\makebox(0,0)[bl]{\(E\)}}

\put(64,20){\circle*{2}}

\put(64,15){\makebox(0,0)[bl]{\(E_{2}\)}}


\put(72,20){\line(0,1){05}}

\put(74,20){\makebox(0,0)[bl]{\(K\)}}

\put(72,20){\circle*{2}}

\put(72,15){\makebox(0,0)[bl]{\(K_{1}\)}}

\end{picture}
\end{center}

\begin{center}
\begin{picture}(77,31)
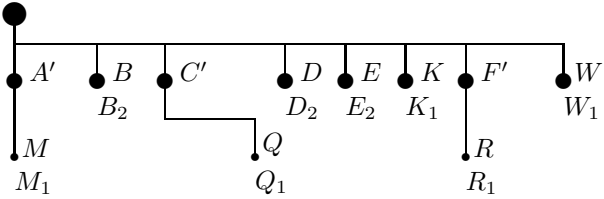
\put(01,0){\makebox(0,0)[bl] {Fig. 9. Forecast
~\(\overline{\Phi}\)~ (multiple choice problem)}}

\put(03,27){\makebox(0,0)[bl]{
}}

\put(00,29){\circle*{3}}

\put(00,25){\line(0,1){04}}

\put(00,25){\line(1,0){73}}


\put(00,20){\line(0,1){05}}

\put(02,20){\makebox(0,0)[bl]{\(A'\)}}

\put(00,20){\circle*{2}}

\put(00,15){\line(0,1){05}}


\put(00,10){\line(0,1){05}}


\put(00,10){\circle*{1.2}}


\put(01,10){\makebox(0,0)[bl]{\(M\)}}

\put(00,05){\makebox(0,0)[bl]{\(M_{1}\)}}



\put(11,20){\line(0,1){05}}

\put(13,20){\makebox(0,0)[bl]{\(B\)}}

\put(11,20){\circle*{2}}

\put(11,15){\makebox(0,0)[bl]{\(B_{2}\)}}







\put(20,20){\line(0,1){05}}


\put(22,20){\makebox(0,0)[bl]{\(C'\)}}

\put(20,20){\circle*{2}}

\put(20,15){\line(0,1){05}}


\put(20,15){\line(1,0){12}}

\put(32,10){\line(0,1){05}}


\put(32,10){\circle*{1.2}}


\put(33,10){\makebox(0,0)[bl]{\(Q\)}}



\put(32,05){\makebox(0,0)[bl]{\(Q_{1}\)}}



\put(36,20){\line(0,1){05}}

\put(38,20){\makebox(0,0)[bl]{\(D\)}}

\put(36,20){\circle*{2}}

\put(36,15){\makebox(0,0)[bl]{\(D_{2}\)}}


\put(44,20){\line(0,1){05}}

\put(46,20){\makebox(0,0)[bl]{\(E\)}}

\put(44,20){\circle*{2}}

\put(44,15){\makebox(0,0)[bl]{\(E_{2}\)}}


\put(52,20){\line(0,1){05}}

\put(54,20){\makebox(0,0)[bl]{\(K\)}}

\put(52,20){\circle*{2}}

\put(52,15){\makebox(0,0)[bl]{\(K_{1}\)}}


\put(60,20){\line(0,1){05}}

\put(62,20){\makebox(0,0)[bl]{\(F'\)}}

\put(60,20){\circle*{2}}

\put(60,15){\line(0,1){05}}

\put(60,10){\line(0,1){05}}

\put(60,10){\circle*{1.2}}

\put(61,10){\makebox(0,0)[bl]{\(R\)}}

\put(60,05){\makebox(0,0)[bl]{\(R_{1}\)}}







\put(73,20){\line(0,1){05}}

\put(74.5,20){\makebox(0,0)[bl]{\(W\)}}

\put(73,20){\circle*{2}}

\put(73,15){\makebox(0,0)[bl]{\(W_{1}\)}}

\end{picture}
\end{center}


 Three obtained solutions (forecasts) can be analyzed:

 (1) direct expert based forecast
   ~\( \widetilde{S_{4}}\)~ (Fig. 6), i.e.,
  corresponding group of improvement operations:~
 \( \widetilde{\Phi} =
 \{ \Phi_{1},\Phi_{2},\Phi_{3},\Phi_{5},\Phi_{7},\Phi_{9},\Phi_{10},\Phi_{13},
 \Phi_{14},\Phi_{15},\Phi_{16},\Phi_{17} \}  \);
 (2) two computation-based forecasts:
%
  ~\( \widehat{\Phi}\) (Fig. 8) and
 ~\( \overline{\Phi} \) (Fig. 9).

 Here the following integrated estimates of the solutions
 are examined (a simplified case):
 ~(i) total  ``profit'':
 ~\( \sum_{\forall \Phi_{i} \in  \widetilde{\Phi}}  ~c_{i} \),
  ~\( \sum_{\forall \Phi_{i} \in  \widehat{\Phi}}  ~c_{i} \),
 ~\( \sum_{\forall \Phi_{i} \in  \overline{\Phi}}  ~c_{i} \);
 ~(ii) total required resource:
 ~\( \sum_{\forall \Phi_{i} \in  \widetilde{\Phi}}  ~a_{i} \),
  ~\( \sum_{\forall \Phi_{i} \in  \widehat{\Phi}}  ~a_{i} \),
 ~\( \sum_{\forall \Phi_{i} \in  \overline{\Phi}}  ~a_{i} \).
 The solutions belong to a layer of Pareto-efficient solutions
 (Fig. 10).

\begin{center}
\begin{picture}(75,64)
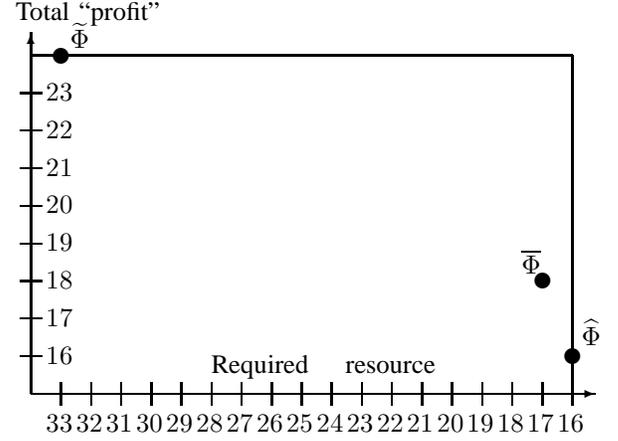
\put(10,00){\makebox(0,0)[bl] {Fig. 10. Comparison of three
 forecasts}}


\put(6,56){\makebox(0,0)[bl]{ \( \widetilde{\Phi} \) }}
\put(6,55){\circle*{2}}

\put(66,26){\makebox(0,0)[bl]{ \( \overline{\Phi}  \) }}
\put(70,25){\circle*{2}}

\put(74,16.5){\makebox(0,0)[bl]{ \( \widehat{\Phi}  \) }}
\put(74,15){\circle*{2}}


\put(0,59){\makebox(0,0)[bl]{Total ``profit''}}


\put(02,10){\vector(0,1){48}}

\put(0.5,15){\line(1,0){03}} \put(0.5,20){\line(1,0){03}}
\put(0.5,25){\line(1,0){03}} \put(0.5,30){\line(1,0){03}}
\put(0.5,35){\line(1,0){03}} \put(0.5,40){\line(1,0){03}}
\put(0.5,45){\line(1,0){03}} \put(0.5,50){\line(1,0){03}}

\put(4,49){\makebox(0,0)[bl]{\(23\)}}
\put(4,44){\makebox(0,0)[bl]{\(22\)}}
\put(4,39){\makebox(0,0)[bl]{\(21\)}}
\put(4,34){\makebox(0,0)[bl]{\(20\)}}
\put(4,29){\makebox(0,0)[bl]{\(19\)}}
\put(4,24){\makebox(0,0)[bl]{\(18\)}}
\put(4,19){\makebox(0,0)[bl]{\(17\)}}
\put(4,14){\makebox(0,0)[bl]{\(16\)}}


\put(02,55){\line(1,0){72}}

\put(74,10){\line(0,1){45}}


\put(26,12){\makebox(0,0)[bl]{Required ~~ resource}}

\put(02,10){\vector(1,0){75}}

\put(6,8.5){\line(0,1){03}} \put(10,8.5){\line(0,1){03}}
\put(14,8.5){\line(0,1){03}} \put(18,8.5){\line(0,1){03}}
\put(22,8.5){\line(0,1){03}} \put(26,8.5){\line(0,1){03}}
\put(30,8.5){\line(0,1){03}} \put(34,8.5){\line(0,1){03}}
\put(38,8.5){\line(0,1){03}} \put(42,8.5){\line(0,1){03}}
\put(46,8.5){\line(0,1){03}} \put(50,8.5){\line(0,1){03}}
\put(54,8.5){\line(0,1){03}} \put(58,8.5){\line(0,1){03}}
\put(62,8.5){\line(0,1){03}} \put(66,8.5){\line(0,1){03}}
\put(70,8.5){\line(0,1){03}} \put(74,8.5){\line(0,1){03}}

\put(04,05){\makebox(0,0)[bl]{\(33\)}}
\put(08,05){\makebox(0,0)[bl]{\(32\)}}
\put(12,05){\makebox(0,0)[bl]{\(31\)}}
\put(16,05){\makebox(0,0)[bl]{\(30\)}}
\put(20,05){\makebox(0,0)[bl]{\(29\)}}
\put(24,05){\makebox(0,0)[bl]{\(28\)}}
\put(28,05){\makebox(0,0)[bl]{\(27\)}}
\put(32,05){\makebox(0,0)[bl]{\(26\)}}
\put(36,05){\makebox(0,0)[bl]{\(25\)}}
\put(40,05){\makebox(0,0)[bl]{\(24\)}}
\put(44,05){\makebox(0,0)[bl]{\(23\)}}
\put(48,05){\makebox(0,0)[bl]{\(22\)}}
\put(52,05){\makebox(0,0)[bl]{\(21\)}}
\put(56,05){\makebox(0,0)[bl]{\(20\)}}
\put(60,05){\makebox(0,0)[bl]{\(19\)}}
\put(64,05){\makebox(0,0)[bl]{\(18\)}}
\put(68,05){\makebox(0,0)[bl]{\(17\)}}
\put(72,05){\makebox(0,0)[bl]{\(16\)}}

\end{picture}
\end{center}

\section{Aggregation of Preliminary Forecasts}

 Two aggregation  strategies
 for aggregation of three preliminary forecasts
  (\(\widetilde{\Phi}\),
 \(\widehat{\Phi}\),
 \(\overline{\Phi}\))
 are considered \cite{lev11agg}:
 (i) extension (addition) strategy I
 (i.e., a ``kernel'' of a substructure of the initial solutions is extended
 by addition of some additional elements),
 (ii) compression (deletion) strategy II
 (i.e.,  a superstructure of the initial solutions is compressed by deletion of some its elements).
 Fig. 11 illustrates the substructure and superstructure
 (as sets of change operations):

 (i) substructure:~
 \( \widetilde{\Phi} \bigcap \overline{\Phi} \bigcap \widehat{\Phi}
 =
  \{ \Phi_{2},\Phi_{5},\Phi_{6} \}\).

 (ii) superstructure:~
  \( \widetilde{\Phi} \bigcup \overline{\Phi} \bigcup \widehat{\Phi}
 =
  \{ \Phi_{1},\Phi_{2},\Phi_{3},\Phi_{4},\Phi_{5},\Phi_{6},
 \)

 \(\Phi_{7},\Phi_{8},\Phi_{9},\Phi_{10},\Phi_{11},
 \Phi_{13},\Phi_{14},\Phi_{15},\Phi_{16},\Phi_{17} \}\).

\begin{center}
\begin{picture}(58,28)
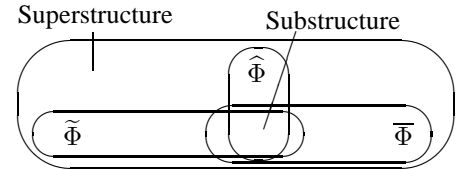
\put(00,0){\makebox(0,0)[bl]{Fig. 11. Substructure and
superstructure}}


\put(20,09){\oval(36,6)}

\put(06,7.5){\makebox(0,8)[bl]{\(\widetilde{\Phi}\)}}


\put(40,09){\oval(30,7.6)}

\put(50,7.5){\makebox(0,8)[bl]{\(\overline{\Phi}\)}}


\put(32,13){\oval(08,15)}

\put(30.5,16){\makebox(0,8)[bl]{\(\widehat{\Phi}\)}}


\put(29,13){\oval(58,17)}

\put(00,23){\makebox(0,8)[bl]{Superstructure}}
 \put(10,22.5){\line(0,-1){5}}

\put(33,23){\makebox(0,8)[bl]{Substructure}}
 \put(37,22.6){\line(-1,-3){4.3}}

\end{picture}
\end{center}

 A list of addition operations (for strategy I)
 is presented in Table 3
 (operations and their estimates correspond
 to Table 2).

\begin{center}
\begin{picture}(87,47)

\put(07,43){\makebox(0,0)[bl] {Table 3. Addition operations
 (estimates,
 priorities)}}


\put(0,0){\line(1,0){87}} \put(0,27){\line(1,0){87}}
\put(0,41){\line(1,0){87}}

\put(0,0){\line(0,1){41}} \put(05,0){\line(0,1){41}}
\put(20,0){\line(0,1){41}} \put(28,0){\line(0,1){41}}
\put(76,0){\line(0,1){41}} \put(87,0){\line(0,1){41}}

\put(35,27){\line(0,1){14}} \put(40,27){\line(0,1){14}}
\put(46,27){\line(0,1){14}} \put(52,27){\line(0,1){14}}
\put(58,27){\line(0,1){14}} \put(64,27){\line(0,1){14}}
\put(70,27){\line(0,1){14}}

\put(2,36.5){\makebox(0,0)[bl]{\(i\)}}

\put(20.5,36.5){\makebox(0,0)[bl]{Vari-}}
\put(20.5,32.5){\makebox(0,0)[bl]{able}}

\put(5.5,36){\makebox(0,0)[bl]{Impro-}}
\put(5.5,32.6){\makebox(0,0)[bl]{vement}}
\put(5.5,28.5){\makebox(0,0)[bl]{operation}}

\put(29,36){\makebox(0,0)[bl]{\(\Upsilon_{1}\)}}
\put(35,36){\makebox(0,0)[bl]{\(\Upsilon_{2}\)}}
\put(41,36){\makebox(0,0)[bl]{\(\Upsilon_{3}\)}}
\put(47,36){\makebox(0,0)[bl]{\(\Upsilon_{4}\)}}
\put(53,36){\makebox(0,0)[bl]{\(\Upsilon_{5}\)}}
\put(59,36){\makebox(0,0)[bl]{\(\Upsilon_{6}\)}}
\put(65,36){\makebox(0,0)[bl]{\(\Upsilon_{7}\)}}
\put(71,36){\makebox(0,0)[bl]{\(\Upsilon_{8}\)}}

\put(76.5,36.5){\makebox(0,0)[bl]{Priori-}}
\put(76.5,32.5){\makebox(0,0)[bl]{ties  \(r_{i}\)}}



\put(1.6,22){\makebox(0,0)[bl]{\(1\)}}

\put(6,21.5){\makebox(0,0)[bl]{\(J_{1}\)(\(\Phi_{10}\))}}

\put(22,21.5){\makebox(0,0)[bl]{\(x_{1}\)}}

\put(1.6,18){\makebox(0,0)[bl]{\(2\)}}

\put(6,17.5){\makebox(0,0)[bl]{\(B_{1}\)(\(\Phi_{13}\))}}

\put(22,17.5){\makebox(0,0)[bl]{\(x_{2}\)}}

\put(1.6,14){\makebox(0,0)[bl]{\(3\)}}

\put(6,13.5){\makebox(0,0)[bl]{\(U_{1}\&U_{2}\)
 }}

\put(6,09.5){\makebox(0,0)[bl]{(\(\Phi_{17}\))}}

\put(22,013.5){\makebox(0,0)[bl]{\(x_{3}\)}}

\put(1.6,6){\makebox(0,0)[bl]{\(4\)}}

\put(6,5.5){\makebox(0,0)[bl]{\(L_{1} \)(\(\Phi_{3}\))}}

\put(22,5.5){\makebox(0,0)[bl]{\(x_{4}\)}}

\put(1.6,02){\makebox(0,0)[bl]{\(5\)}}

\put(6,01.5){\makebox(0,0)[bl]{\(W_{1}\)(\(\Phi_{15}\))}}

\put(22,01.5){\makebox(0,0)[bl]{\(x_{5}\)}}


\put(30,22){\makebox(0,0)[bl]{\(2\)}}
\put(36,22){\makebox(0,0)[bl]{\(3\)}}

\put(42,22){\makebox(0,0)[bl]{\(4\)}}

\put(48,22){\makebox(0,0)[bl]{\(4\)}}
\put(54,22){\makebox(0,0)[bl]{\(3\)}}

\put(60,22){\makebox(0,0)[bl]{\(4\)}}
\put(66,22){\makebox(0,0)[bl]{\(4\)}}
\put(72,22){\makebox(0,0)[bl]{\(3\)}}

\put(80.5,22){\makebox(0,0)[bl]{\(1\)}}

\put(30,18){\makebox(0,0)[bl]{\(3\)}}
\put(36,18){\makebox(0,0)[bl]{\(3\)}}

\put(42,18){\makebox(0,0)[bl]{\(3\)}}

\put(48,18){\makebox(0,0)[bl]{\(4\)}}
\put(54,18){\makebox(0,0)[bl]{\(3\)}}

\put(60,18){\makebox(0,0)[bl]{\(3\)}}
\put(66,18){\makebox(0,0)[bl]{\(4\)}}
\put(72,18){\makebox(0,0)[bl]{\(3\)}}

\put(80.5,18){\makebox(0,0)[bl]{\(3\)}}

\put(30,14){\makebox(0,0)[bl]{\(3\)}}
\put(36,14){\makebox(0,0)[bl]{\(3\)}}

\put(42,14){\makebox(0,0)[bl]{\(4\)}}

\put(48,14){\makebox(0,0)[bl]{\(3\)}}
\put(54,14){\makebox(0,0)[bl]{\(4\)}}

\put(60,14){\makebox(0,0)[bl]{\(3\)}}
\put(66,14){\makebox(0,0)[bl]{\(3\)}}
\put(72,14){\makebox(0,0)[bl]{\(3\)}}

\put(80.5,14){\makebox(0,0)[bl]{\(3\)}}

\put(30,06){\makebox(0,0)[bl]{\(3\)}}
\put(36,06){\makebox(0,0)[bl]{\(4\)}}

\put(42,06){\makebox(0,0)[bl]{\(3\)}}

\put(48,06){\makebox(0,0)[bl]{\(4\)}}
\put(54,06){\makebox(0,0)[bl]{\(3\)}}

\put(60,06){\makebox(0,0)[bl]{\(5\)}}
\put(66,06){\makebox(0,0)[bl]{\(3\)}}
\put(72,06){\makebox(0,0)[bl]{\(4\)}}

\put(80.5,06){\makebox(0,0)[bl]{\(1\)}}

\put(30,02){\makebox(0,0)[bl]{\(3\)}}
\put(36,02){\makebox(0,0)[bl]{\(3\)}}

\put(42,02){\makebox(0,0)[bl]{\(2\)}}

\put(48,02){\makebox(0,0)[bl]{\(3\)}}
\put(54,02){\makebox(0,0)[bl]{\(4\)}}

\put(60,02){\makebox(0,0)[bl]{\(3\)}}
\put(66,02){\makebox(0,0)[bl]{\(3\)}}
\put(72,02){\makebox(0,0)[bl]{\(5\)}}

\put(80.5,02){\makebox(0,0)[bl]{\(2\)}}

\end{picture}
\end{center}

  The addition problem (simplified knapsack problem) is:
 \[\max \sum_{i=1}^{5} c_{i} x_{i} ~~~
%
  s.t. \sum_{i=1}^{5}  ~a_{i} x_{i} \leq b, ~~x_{i} \in \{0,1\}.\]
 Cost estimates (by criterion \(\Upsilon_{1}\))
 are used as
 \(\{a_{i}\}\),
  priorities \(\{r_{i}\}\) are used as (transform to)
  \(\{c_{i}\}\),
  and
   \(b = 8.00\).
%
 A resultant solution for strategy I is depicted in Fig. 12
 (\(x_{1} = 1\), \(x_{2} = 1\), \(x_{3} = 0\), \(x_{4} = 0\), \(x_{5} = 1\))
 (a simple greedy algorithm was used).
\begin{center}
\begin{picture}(84,33)

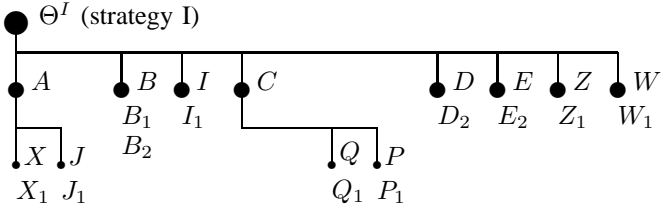
\put(9,00){\makebox(0,0)[bl] {Fig. 12. Aggregated
 forecast
 \(\Theta^{I}\) (strategy I) }}

\put(03,28){\makebox(0,0)[bl]{\(\Theta^{I}\) (strategy I)
}}

\put(00,29){\circle*{3}}

\put(00,25){\line(0,1){04}}


\put(00,25){\line(1,0){80}}


\put(00,20){\line(0,1){05}}

\put(02,20){\makebox(0,0)[bl]{\(A\)}}

\put(00,20){\circle*{2}}

\put(00,15){\line(0,1){05}}

\put(00,15){\line(1,0){6}}

\put(00,10){\line(0,1){05}}

\put(06,10){\line(0,1){05}}

\put(00,10){\circle*{1.2}}

\put(06,10){\circle*{1.2}}

\put(01,10){\makebox(0,0)[bl]{\(X\)}}
\put(07,10){\makebox(0,0)[bl]{\(J\)}}

\put(00,05){\makebox(0,0)[bl]{\(X_{1}\)}}

\put(06,05){\makebox(0,0)[bl]{\(J_{1}\)}}


\put(14,20){\line(0,1){05}}

\put(16,20){\makebox(0,0)[bl]{\(B\)}}

\put(14,20){\circle*{2}}

\put(14,15){\makebox(0,0)[bl]{\(B_{1}\)}}
\put(14,11){\makebox(0,0)[bl]{\(B_{2}\)}}


\put(22,20){\line(0,1){05}}

\put(24,20){\makebox(0,0)[bl]{\(I\)}}

\put(22,20){\circle*{2}}

\put(22,15){\makebox(0,0)[bl]{\(I_{1}\)}}


\put(30,20){\line(0,1){05}}


\put(32,20){\makebox(0,0)[bl]{\(C\)}}

\put(30,20){\circle*{2}}

\put(30,15){\line(0,1){05}}

\put(30,15){\line(1,0){18}}

\put(30,15){\line(1,0){12}}


\put(42,10){\line(0,1){05}}

\put(48,10){\line(0,1){05}}


\put(42,10){\circle*{1.2}}

\put(48,10){\circle*{1.2}}

\put(43,10){\makebox(0,0)[bl]{\(Q\)}}
\put(49,10){\makebox(0,0)[bl]{\(P\)}}


\put(42,05){\makebox(0,0)[bl]{\(Q_{1}\)}}

\put(48,05){\makebox(0,0)[bl]{\(P_{1}\)}}


\put(56,20){\line(0,1){05}}

\put(58,20){\makebox(0,0)[bl]{\(D\)}}

\put(56,20){\circle*{2}}

\put(56,15){\makebox(0,0)[bl]{\(D_{2}\)}}


\put(64,20){\line(0,1){05}}

\put(66,20){\makebox(0,0)[bl]{\(E\)}}

\put(64,20){\circle*{2}}

\put(64,15){\makebox(0,0)[bl]{\(E_{2}\)}}


\put(72,20){\line(0,1){05}}

\put(74,20){\makebox(0,0)[bl]{\(Z\)}}

\put(72,20){\circle*{2}}

\put(72,15){\makebox(0,0)[bl]{\(Z_{1}\)}}













\put(80,20){\line(0,1){05}}

\put(82,20){\makebox(0,0)[bl]{\(W\)}}

\put(80,20){\circle*{2}}

\put(80,15){\makebox(0,0)[bl]{\(W_{1}\)}}

\end{picture}
\end{center}

  A list of deletion operations (for strategy II)
 is presented in Table 4.

\begin{center}
\begin{picture}(87,52)

\put(07.5,47){\makebox(0,0)[bl]{Table 4. Deletion operations
 (estimates, priorities)}}


\put(0,0){\line(1,0){87}} \put(0,35){\line(1,0){87}}
\put(0,45){\line(1,0){87}}

\put(0,0){\line(0,1){45}} \put(05,0){\line(0,1){45}}
\put(19,0){\line(0,1){45}} \put(28,0){\line(0,1){45}}
\put(76,0){\line(0,1){45}} \put(87,0){\line(0,1){45}}

\put(34,35){\line(0,1){10}} \put(40,35){\line(0,1){10}}
\put(46,35){\line(0,1){10}} \put(52,35){\line(0,1){10}}
\put(58,35){\line(0,1){10}} \put(64,35){\line(0,1){10}}
\put(70,35){\line(0,1){10}}

\put(2,40.5){\makebox(0,0)[bl]{\(i\)}}

\put(5.5,40.5){\makebox(0,0)[bl]{Deletion}}
\put(5.5,36.5){\makebox(0,0)[bl]{operation}}

\put(19.5,40.5){\makebox(0,0)[bl]{Vari-}}
\put(19.5,36.5){\makebox(0,0)[bl]{able}}

\put(29,40){\makebox(0,0)[bl]{\(\Upsilon_{1}\)}}
\put(35,40){\makebox(0,0)[bl]{\(\Upsilon_{2}\)}}
\put(41,40){\makebox(0,0)[bl]{\(\Upsilon_{3}\)}}
\put(47,40){\makebox(0,0)[bl]{\(\Upsilon_{4}\)}}

\put(53,40){\makebox(0,0)[bl]{\(\Upsilon_{5}\)}}
\put(59,40){\makebox(0,0)[bl]{\(\Upsilon_{6}\)}}
\put(65,40){\makebox(0,0)[bl]{\(\Upsilon_{7}\)}}
\put(71,40){\makebox(0,0)[bl]{\(\Upsilon_{8}\)}}

\put(76.5,41){\makebox(0,0)[bl]{Priori-}}
\put(76.5,37){\makebox(0,0)[bl]{ties  \(r_{i}\)}}








\put(1.6,30){\makebox(0,0)[bl]{\(1\)}}
\put(6,29.5){\makebox(0,0)[bl]{\(J_{1}\)(\(\Phi_{10}\))}}
\put(22,30){\makebox(0,0)[bl]{\(x_{1}\)}}

\put(1.6,26){\makebox(0,0)[bl]{\(2\)}}
\put(6,25.5){\makebox(0,0)[bl]{\(B_{1}\)(\(\Phi_{13}\))}}
\put(22,26){\makebox(0,0)[bl]{\(x_{2}\)}}

\put(1.6,22){\makebox(0,0)[bl]{\(3\)}}
\put(6,21.5){\makebox(0,0)[bl]{\(Q_{1}\)(\(\Phi_{7})\)}}
\put(22,22){\makebox(0,0)[bl]{\(x_{3}\)}}

\put(1.6,18){\makebox(0,0)[bl]{\(4\)}}
\put(6,17.5){\makebox(0,0)[bl]{\(P_{1}\)(\(\Phi_{8}\))}}
\put(22,18){\makebox(0,0)[bl]{\(x_{4}\)}}

\put(1.6,14){\makebox(0,0)[bl]{\(5\)}}
\put(6,13.5){\makebox(0,0)[bl]{\(E_{3}\)(\(\Phi_{14}\))}}
\put(22,14){\makebox(0,0)[bl]{\(x_{5}\)}}

\put(1.6,10){\makebox(0,0)[bl]{\(6\)}}
\put(6,09.5){\makebox(0,0)[bl]{\(L_{1}\)(\(\Phi_{3}\))}}
\put(22,10){\makebox(0,0)[bl]{\(x_{6}\)}}

\put(1.6,6){\makebox(0,0)[bl]{\(7\)}}
\put(6,05.5){\makebox(0,0)[bl]{\(U_{1}\&U_{2}\)
}}

\put(6,01.5){\makebox(0,0)[bl]{(\(\Phi_{17}\))}}

\put(22,06){\makebox(0,0)[bl]{\(x_{7}\)}}


\put(30,30){\makebox(0,0)[bl]{\(2\)}}
\put(36,30){\makebox(0,0)[bl]{\(3\)}}

\put(42,30){\makebox(0,0)[bl]{\(4\)}}

\put(48,30){\makebox(0,0)[bl]{\(4\)}}
\put(54,30){\makebox(0,0)[bl]{\(3\)}}

\put(60,30){\makebox(0,0)[bl]{\(4\)}}
\put(66,30){\makebox(0,0)[bl]{\(4\)}}
\put(72,30){\makebox(0,0)[bl]{\(3\)}}

\put(80.5,30){\makebox(0,0)[bl]{\(1\)}}

\put(30,26){\makebox(0,0)[bl]{\(3\)}}
\put(36,26){\makebox(0,0)[bl]{\(3\)}}

\put(42,26){\makebox(0,0)[bl]{\(3\)}}
\put(48,26){\makebox(0,0)[bl]{\(4\)}}

\put(54,26){\makebox(0,0)[bl]{\(3\)}}

\put(60,26){\makebox(0,0)[bl]{\(3\)}}
\put(66,26){\makebox(0,0)[bl]{\(4\)}}
\put(72,26){\makebox(0,0)[bl]{\(3\)}}

\put(80.5,26){\makebox(0,0)[bl]{\(3\)}}

\put(30,22){\makebox(0,0)[bl]{\(3\)}}
\put(36,22){\makebox(0,0)[bl]{\(4\)}}

\put(42,22){\makebox(0,0)[bl]{\(3\)}}

\put(48,22){\makebox(0,0)[bl]{\(3\)}}
\put(54,22){\makebox(0,0)[bl]{\(3\)}}

\put(60,22){\makebox(0,0)[bl]{\(2\)}}
\put(66,22){\makebox(0,0)[bl]{\(3\)}}
\put(72,22){\makebox(0,0)[bl]{\(3\)}}

\put(80.5,22){\makebox(0,0)[bl]{\(4\)}}

\put(30,18){\makebox(0,0)[bl]{\(3\)}}
\put(36,18){\makebox(0,0)[bl]{\(4\)}}

\put(42,18){\makebox(0,0)[bl]{\(4\)}}

\put(48,18){\makebox(0,0)[bl]{\(4\)}}
\put(54,18){\makebox(0,0)[bl]{\(3\)}}

\put(60,18){\makebox(0,0)[bl]{\(4\)}}
\put(66,18){\makebox(0,0)[bl]{\(3\)}}
\put(72,18){\makebox(0,0)[bl]{\(3\)}}

\put(80.5,18){\makebox(0,0)[bl]{\(2\)}}

\put(30,14){\makebox(0,0)[bl]{\(3\)}}
\put(36,14){\makebox(0,0)[bl]{\(3\)}}

\put(42,14){\makebox(0,0)[bl]{\(3\)}}

\put(48,14){\makebox(0,0)[bl]{\(3\)}}
\put(54,14){\makebox(0,0)[bl]{\(5\)}}

\put(60,14){\makebox(0,0)[bl]{\(3\)}}
\put(66,14){\makebox(0,0)[bl]{\(3\)}}
\put(72,14){\makebox(0,0)[bl]{\(3\)}}

\put(80.5,14){\makebox(0,0)[bl]{\(2\)}}

\put(30,10){\makebox(0,0)[bl]{\(3\)}}
\put(36,10){\makebox(0,0)[bl]{\(4\)}}

\put(42,10){\makebox(0,0)[bl]{\(3\)}}

\put(48,10){\makebox(0,0)[bl]{\(4\)}}
\put(54,10){\makebox(0,0)[bl]{\(3\)}}

\put(60,10){\makebox(0,0)[bl]{\(5\)}}
\put(66,10){\makebox(0,0)[bl]{\(3\)}}
\put(72,10){\makebox(0,0)[bl]{\(4\)}}

\put(80.5,10){\makebox(0,0)[bl]{\(1\)}}

\put(30,6){\makebox(0,0)[bl]{\(3\)}}
\put(36,6){\makebox(0,0)[bl]{\(3\)}}

\put(42,6){\makebox(0,0)[bl]{\(4\)}}

\put(48,6){\makebox(0,0)[bl]{\(3\)}}
\put(54,6){\makebox(0,0)[bl]{\(4\)}}

\put(60,6){\makebox(0,0)[bl]{\(3\)}}
\put(66,6){\makebox(0,0)[bl]{\(3\)}}
\put(72,6){\makebox(0,0)[bl]{\(3\)}}

\put(80.5,6){\makebox(0,0)[bl]{\(3\)}}

\end{picture}
\end{center}

  The deletion problem (knapsack problem with minimization
 of the objective function) is:
 \[\min \sum_{i=1}^{7} c_{i} x_{i} ~~~
%
  s.t. \sum_{i=1}^{7} ~ a_{i} x_{i} \geq b, ~~x_{i} \in \{0,1\}.\]
 Cost estimates are (by criterion \(\Upsilon_{1}\)) used as
 \(\{a_{i}\}\),
  priorities \(\{r_{i}\}\) are used as (transform to)
  \(\{c_{i}\}\),
  and \(b=8.00\).
%
%
 A resultant solution based on strategy II is depicted in Fig. 13
 (\(x_{1} = 0\), \(x_{2} = 1\), \(x_{3} = 1\), \(x_{4} = 1\),
 \(x_{5} = 0\), \(x_{6} = 0\), \(x_{7} = 1\))
 (a simple greedy algorithm was used).

\begin{center}
\begin{picture}(83,37)

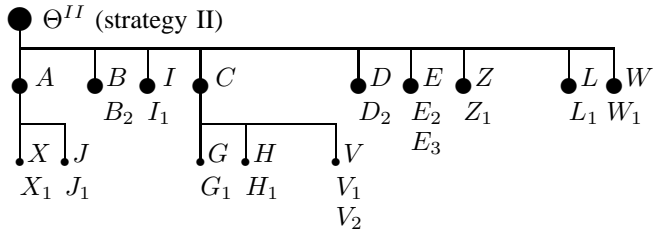
\put(07,00){\makebox(0,0)[bl] {Fig. 13. Aggregated
 forecast
 \(\Theta^{II}\) (strategy II) }}

\put(03,32){\makebox(0,0)[bl]{\(\Theta^{II}\) (strategy II)
 }}

\put(00,34){\circle*{3}}

\put(00,30){\line(0,1){04}}

\put(00,30){\line(1,0){79}}


\put(00,25){\line(0,1){05}}

\put(02,25){\makebox(0,0)[bl]{\(A\)}}

\put(00,25){\circle*{2}}

\put(00,20){\line(0,1){05}}

\put(00,20){\line(1,0){6}}

\put(00,15){\line(0,1){05}} \put(06,15){\line(0,1){05}}

\put(00,15){\circle*{1.2}} \put(06,15){\circle*{1.2}}

\put(01,15){\makebox(0,0)[bl]{\(X\)}}
\put(07,15){\makebox(0,0)[bl]{\(J\)}}

\put(00,10){\makebox(0,0)[bl]{\(X_{1}\)}}

\put(06,10){\makebox(0,0)[bl]{\(J_{1}\)}}


\put(10,25){\line(0,1){05}}

\put(11.5,25){\makebox(0,0)[bl]{\(B\)}}

\put(10,25){\circle*{2}}

\put(11,20){\makebox(0,0)[bl]{\(B_{2}\)}}


\put(17,25){\line(0,1){05}}

\put(19,25){\makebox(0,0)[bl]{\(I\)}}

\put(17,25){\circle*{2}}

\put(17,20){\makebox(0,0)[bl]{\(I_{1}\)}}


\put(24,25){\line(0,1){05}}

\put(26,25){\makebox(0,0)[bl]{\(C\)}}

\put(24,25){\circle*{2}}

\put(24,20){\line(0,1){05}}

\put(24,20){\line(1,0){18}}

\put(24,15){\line(0,1){05}} \put(30,15){\line(0,1){05}}


\put(42,15){\line(0,1){05}}

\put(24,15){\circle*{1.2}} \put(30,15){\circle*{1.2}}


\put(42,15){\circle*{1.2}}

\put(25,15){\makebox(0,0)[bl]{\(G\)}}
\put(31,15){\makebox(0,0)[bl]{\(H\)}}
\put(43,15){\makebox(0,0)[bl]{\(V\)}}

\put(24,10){\makebox(0,0)[bl]{\(G_{1}\)}}
\put(30,10){\makebox(0,0)[bl]{\(H_{1}\)}}
\put(42,10){\makebox(0,0)[bl]{\(V_{1}\)}}
\put(42,06){\makebox(0,0)[bl]{\(V_{2}\)}}


\put(45,25){\line(0,1){05}}

\put(46.5,25){\makebox(0,0)[bl]{\(D\)}}

\put(45,25){\circle*{2}}

\put(45,20){\makebox(0,0)[bl]{\(D_{2}\)}}


\put(52,25){\line(0,1){05}}

\put(53.5,25){\makebox(0,0)[bl]{\(E\)}}

\put(52,25){\circle*{2}}

\put(52,20){\makebox(0,0)[bl]{\(E_{2}\)}}
\put(52,16){\makebox(0,0)[bl]{\(E_{3}\)}}


\put(59,25){\line(0,1){05}}

\put(60.5,25){\makebox(0,0)[bl]{\(Z\)}}

\put(59,25){\circle*{2}}

\put(59,20){\makebox(0,0)[bl]{\(Z_{1}\)}}


\put(73,25){\line(0,1){05}}

\put(74.5,25){\makebox(0,0)[bl]{\(L\)}}

\put(73,25){\circle*{2}}

\put(73,20){\makebox(0,0)[bl]{\(L_{1}\)}}


\put(79,25){\line(0,1){05}}

\put(80.5,25){\makebox(0,0)[bl]{\(W\)}}

\put(79,25){\circle*{2}}

\put(78,20){\makebox(0,0)[bl]{\(W_{1}\)}}

\end{picture}
\end{center}

\section{Conclusion}

 In the paper,
 we have firstly suggested and  described the following:
 (a) a hierarchical modular model for communication protocol ZigBee,
 (b) typical change operations (between protocol generations) and their evaluation,
 (c) a direct expert-based protocol ZigBee/IP (6LoWPAN) 2010,
 (d) two computed protocol forecasts,
 (e) aggregation of the obtained protocol forecasts to build
  two aggregated forecasts.

 It is reasonable to consider the following future research directions:
 {\it 1.} consideration and usage of special comparison analysis
 approaches for  protocol forecasts;
 {\it 2.} examination of other communication protocols
 (and other applied modular systems);
 {\it 3.} usage of fuzzy set based approaches
 and corresponding problems/models;
 {\it 4.} usage of AI-based methods.
 and
 ~{\it 5.} usage of the approaches to combinatorial evolution
 and forecasting in engineering education.

\end{document}